\documentclass[11pt,a4paper]{article}
\pdfoutput=1 

\usepackage{jcappub} 

\usepackage[T1]{fontenc} 

\definecolor{armygreen}{rgb}{0.29, 0.33, 0.13}

\usepackage{url}
\usepackage{graphics}
\newcommand{\gray}{$\gamma$-ray}

\newcommand{\OM}[1]{\textcolor{black}{#1}}
\newcommand{\A}[1]{\textcolor{black}{#1}}
\newcommand{\D}[1]{\textcolor{black}{#1}}
\newcommand{\sv}{\langle \sigma v \rangle}

\newcommand{\DvI}[1]{\textcolor{black}{#1}}
\newcommand{\OMvI}[1]{\textcolor{black}{#1}}
\newcommand{\ajnvI}[1]{\textcolor{black}{#1}}

\newif\iffigure
\figuretrue

\title{\boldmath Constraining dark matter annihilation with HSC Low Surface Brightness Galaxies}

\author[a,1]{Daiki Hashimoto,\note{E-mail address: hashimoto.daiki@f.mbox.nagoya-u.ac.jp}}
\author[b,c]{Oscar Macias,}
\author[d,a]{Atsushi J. Nishizawa,}
\author[e]{Kohei Hayashi,}
\author[f]{Masahiro Takada,}
\author[g]{Masato Shirasaki,}
\author[c,b]{and Shin'ichiro Ando}

\affiliation[a]{Division of particle and astrophysical sciences, Graduate School of Science, Nagoya University, Furocho Chikusa, Nagoya, 464-8602, Aichi, Japan}
\affiliation[b]{Kavli Institute for the Physics and Mathematics of the
Universe (WPI), University of Tokyo, Kashiwa, Chiba 277-8583, Japan}
\affiliation[c]{GRAPPA Institute, University of Amsterdam, 1098 XH
Amsterdam, The Netherlands}
\affiliation[d]{Institute for Advanced Research, Nagoya University, Furocho Chikusa, Nagoya, 464-8602, Aichi, Japan}
\affiliation[e]{Institute for Cosmic Ray Research, The University of Tokyo, Chiba 277-8583, Japan}
\affiliation[f]{Kavli Institute for the Physics and Mathematics of the Universe (WPI), The University of Tokyo Institutes for Advanced Study (UTIAS), The University of Tokyo, 5-1-5 Kashiwanoha, Kashiwa-shi, Chiba, 277-8583, Japan}
\affiliation[g]{National Astronomical Observatory of Japan, Mitaka, Tokyo 181-8588, Japan}

\abstract{
Searches for dark matter annihilation signals have been carried out in a number of target regions such as the Galactic Center and Milky Way dwarf spheroidal galaxies (dSphs), among a few others.
Here we propose low surface brightness galaxies~(LSBGs) as novel targets for the indirect detection of dark matter emission. In particular, LSBGs are known to have very large dark matter contents and 
be less contaminated by extragalactic \gray~sources (e.g., blazars) compared to star forming galaxies.
We report on an analysis that uses eight LSBGs (detected by Subaru Hyper Suprime-Cam survey data) with known redshifts to conduct a search for $\gamma$-ray emission at the positions of these new objects in Fermi Large Area Telescope data. We found no excesses of $\gamma$-ray emission and set constraints on the dark matter annihilation cross-section. 
We exclude (at the 95\% C.L.) dark matter scenarios predicting a cross-section higher than
$\sim 10^{-23}~[\rm cm^3/s]$ for dark matter particles of mass 10 GeV self-annihilating in the $b\bar{b}$ channel.
Although this constraint is weaker than the ones reported in recent studies using other targets, 
we note that in the near future, 
the number of detections of new LSBGs will increase by a few orders of magnitude. 
\OMvI{We forecast that with the use of the full catalog of soon-to-be-detected LSBGs the constraint will reach cross-section sensitivities of $\sim 3\times 10^{-25}~[\rm cm^3/s]$ for dark matter particles with masses $\lesssim 10$ GeV.} 
}

\begin{document}
\maketitle
\flushbottom

\section{Introduction}
\label{sec:intro}
It has been discovered that dark matter~(DM) accounts for $\sim25\%$ of the energy density of the Universe \cite{planck2018+}.
Despite great efforts over several decades DM 
has not been identified yet.
Unravelling the nature of DM is one of the most important subjects in astrophysics.
Weakly Interacting Massive Particles~(WIMPs) have been considered 
one of the best theoretically motivated
candidates for a dominant fraction of DM~(e.g., \citep{2018RPPh...81f6201R, 2019arXiv190407915L}).
They can be produced in thermal equilibrium through interactions with standard model particles in the early Universe,
and therefore some possible processes in which WIMPs self-annihilate into standard model particles can exist \citep{1996PhR...267..195J, 2005PhR...405..279B, 2019arXiv190303616A}.
To explain the present abundance of DM, a total cross-section for such annihilation processes is estimated to be $\sim 3\times 10^{-26}~[\rm cm^3/s]$ for DM masses of order a few GeV (for a more accurate calculation, see \cite{2012PhRvD..86b3506S}).
In these processes, $\gamma$ rays are produced through 
primary or secondary processes~(cascades into unstable states that lead to $\gamma$-ray production).
Although $\gamma$ rays induced by those processes 
are expected to be rare, we can search for DM self-annihilation signals in regions of high DM concentration in the Universe.

Such $\gamma$-rays can be observed by detectors like the Fermi Large Area Telescope~(\emph{Fermi}-LAT) on board the Fermi Gamma-Ray Space Telescope.
\emph{Fermi}-LAT is the 
most sensitive $\gamma$ rays telescope in the energy range of 20~MeV to 1~TeV
~(\citep{2009ApJ...697.1071A, 2012ApJS..203....4A, 2013arXiv1303.3514A, 2018arXiv181011394B}). Thanks to its unprecedented angular resolution and sensitivity, \emph{Fermi}-LAT 
has detected
thousands of $\gamma$-ray sources~\citep{3FGL, 4FGL}.
Close to the Galactic plane the Galactic diffuse emission produced by the interaction of energetic cosmic rays with interstellar gas, ambient photons and magnetic fields dominates. While at high latitudes, the isotropic $\gamma$-ray emission$-$thought to be of extragalactic origin$-$is the most dominant component and have been investigated about its origin (e.g., \citep{1994PAN....57..425K, 1994PAN....57.1268K, 1995PhRvD..52.1828F, 1999JETPL..69..434G, 2008PAN....71..147B, 2011ApJ...733...66I, 2012ApJ...755..164A, 2014PhRvD..90b3514A, 2014ApJ...780...73A, 2015ApJ...799...86A, 2015ApJ...800L..27A, 2015ApJ...807...77K, 2018PhRvL.121x1101A, 2019MNRAS.484.5256H}).


In recent studies, nearby objects like the center of the Milky Way, dwarf spheroidal galaxies~(dSphs), the Local group of galaxies and nearby galaxy clusters have been used as a targets for indirect searches.
For example, several different teams \citep{Zeldovich:1980st, 2009arXiv0910.2998G, 2014GrCo...20...47B, 2013PhRvD..88h3521G, 2014PhRvD..90b3526A, 2015JCAP...03..038C, 2016PDU....12....1D, 2017ApJ...840...43A}, have found an excess $\gamma$-ray emission in the Galactic Center which has been shown to roughly match the expected characteristics of a DM annihilation signal.


Although the origin of the Galactic Center excess has not yet been singled out, recent studies have presented strong evidence that the origin of this signal is related to the stars (e.g., \citep{ 2018NatAs...2..387M, 2018NatAs...2..819B, 2019arXiv190103822M}).
It is true that the Galactic center is where the most luminous annihilating DM emission is expected, however, it is difficult to 
disentangle a DM signal from this region due to large background and foreground uncertainties. 

The authors of \citep{2014PhRvD..89d2001A, 2015PhRvL.115w1301A, 2017ApJ...834..110A, 2018PhRvD..98h3008G, 2018arXiv181206986H, 2018PhRvD..97i5031B} have searched for DM emission in Milky Way dSphs, which are perhaps the most desirable targets to date
because they are relatively close to us and have less astronomical $\gamma$-ray contamination.
In fact, in the case of the Milky Way dSphs, the strongest constraints on the velocity averaged DM cross-section have been provided.

In this work, we propose a new target for probing DM annihilation; low surface brightness galaxies~(LSBG),
which are known to have surface brightness of order ~$>23~{\rm mag/arcmin^2}$ in most cases.
Due to their very low brightness, LSBGs are more difficult to be identified than ordinary galaxies.
Currently, surveys with high sensitivity like Subaru Hyper Suprime-Cam~(HSC), has made it possible to discover nearby LSBGs in large amounts.
The mass-to-light ratio of LSBG is likely to be large (50-100),
thus, those systems are highly dominated by DM \citep{2017MNRAS.470.1512W, 2019MNRAS.483.1754D}.
LSBGs have more massive halos than Milky Way dSphs and present less star-formation, pulsar or supernovae emission than those of ordinary galaxies or galaxy clusters \citep{2019MNRAS.484.4865P}.
Therefore, it is expected that they are mostly quiescent in $\gamma$-rays.

Since LSBGs have a smaller angular size than the point spread function~(PSF) of the \emph{Fermi}-LAT, it is not possible to decompose LSBGs into their substructures. So in practice, LSBGs can be treated as point sources in our $\gamma$-ray analysis. Importantly, since our sample of LSBG are located at high latitudes, the impact of uncertainties in the Galactic diffuse emission is expected to be small. This will naturally allow us to obtain DM limits that can be regarded as being robust to systematic uncertainties in the Galactic diffuse emission model.

In this analysis, we use an LSBG catalog \citep{2018ApJ...857..104G} including $\sim 800$ objects by Subaru HSC observations of $\sim200~{\rm deg^2}$ of the sky.
Unfortunately, we currently have only eight LSBGs with known accurate redshifts, hence here we only employ this small sample of LSBGs.
Using $\gamma$ ray observations taken with the \emph{Fermi}-LAT,
we constrain the DM velocity averaged cross-section for DM masses of order a few GeV.
Next generation telescopes like the Large Synoptic Survey Telescope~(LSST) is expected to discover large amounts of LSBGs due to the large survey areas it can cover and sufficient depths, which will improve our DM constraints significantly. 
In the future, the importance of searching for DM annihilation using LSBGs should increase and it is worthwhile to probe the DM annihilation using LSBGs at present.

This paper is organized as follows.
In Section~\ref{sec:data}, we introduce a sample of HSC LSBGs and $\gamma$-ray observation by \emph{Fermi}-LAT used in our analysis,
In Section~\ref{sec:model}, we derive formulations of the $\gamma$-ray flux induced by the DM annihilation within LSBGs and describe
the method of analysis to probe the annihilation cross-section of DM.
Our results are shown in Section \ref{sec:result} and finally, summary and discussion are described in Section~\ref{sec:summary}.

\section{Data}
\label{sec:data}
\subsection{HSC LSBG sample}
\label{ssec:lsbg}
\begin{figure*}
\begin{center}
\begin{tabular}{cccc}
     \includegraphics[width=0.22\linewidth]{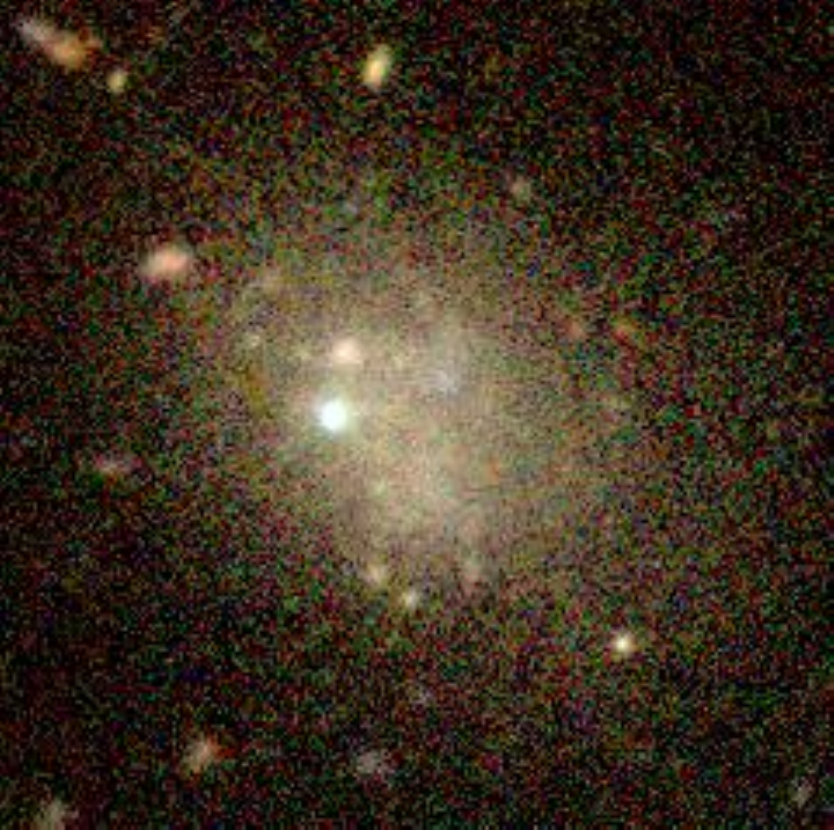}&
     \includegraphics[width=0.22\linewidth]{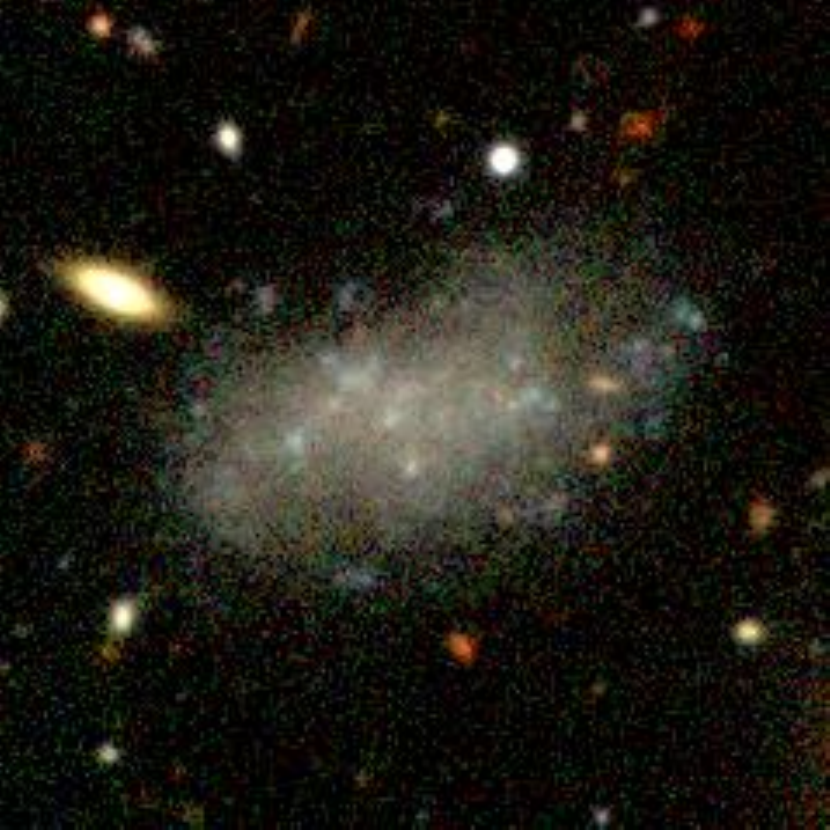}&
     \includegraphics[width=0.22\linewidth]{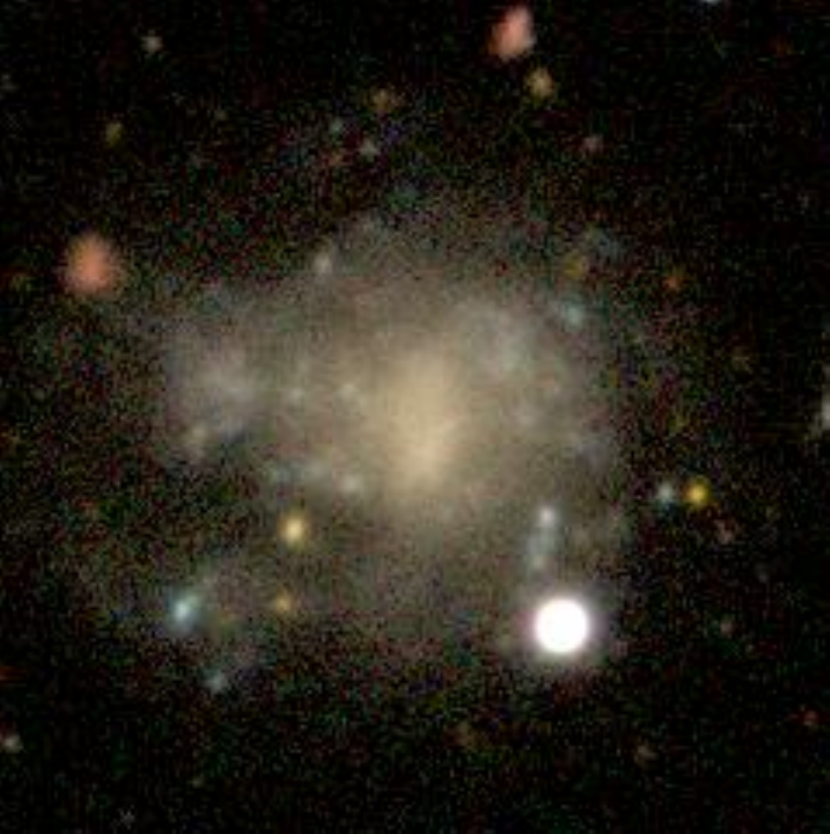}&
     \includegraphics[width=0.22\linewidth]{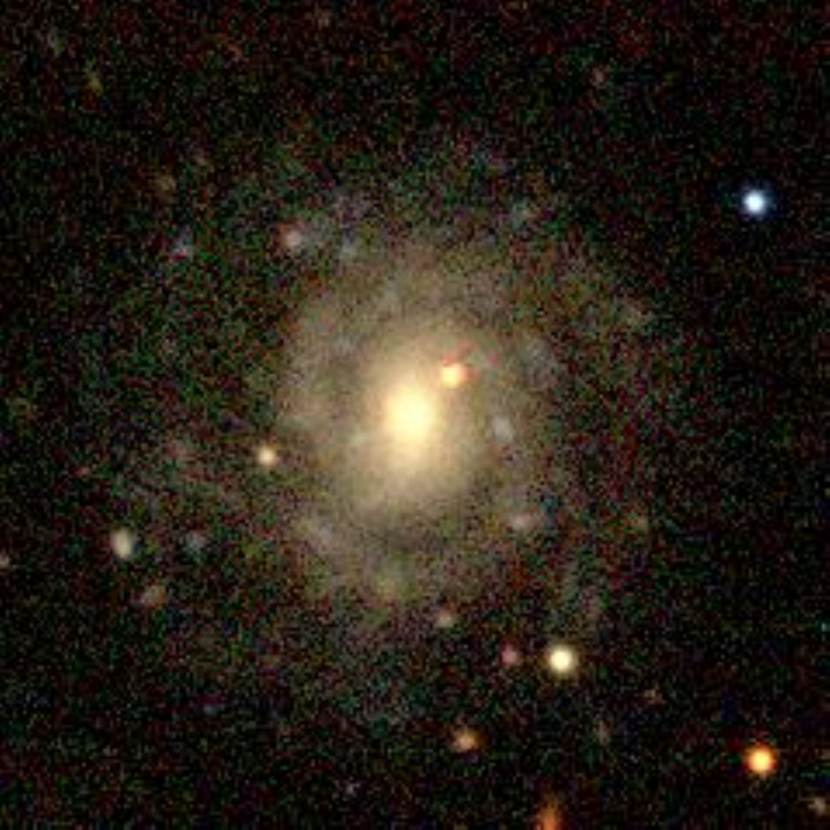}\\
     \includegraphics[width=0.22\linewidth]{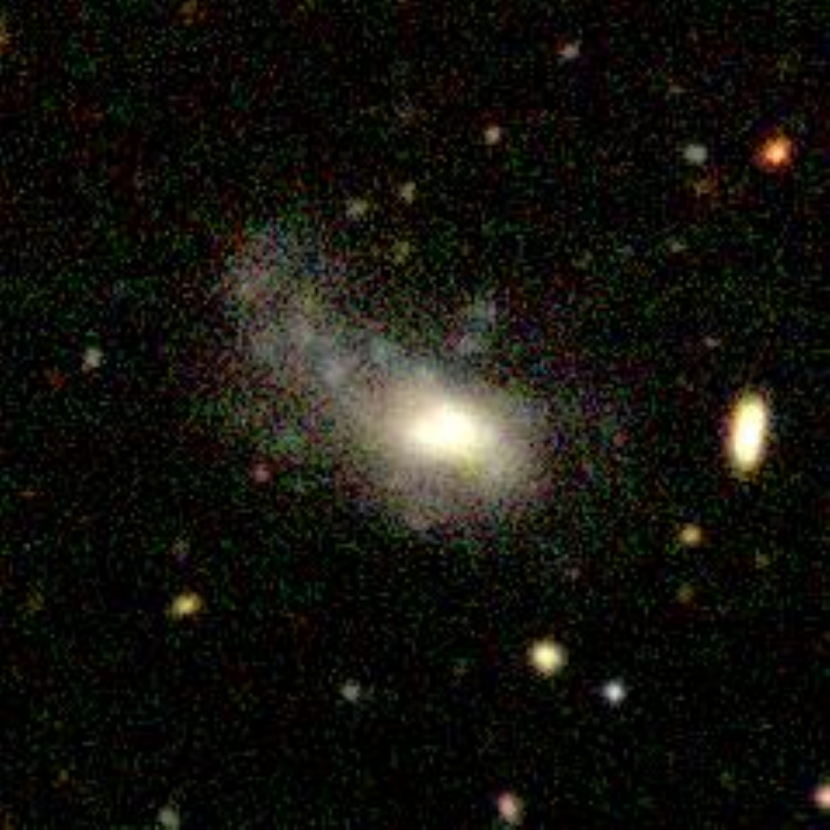}&
     \includegraphics[width=0.22\linewidth]{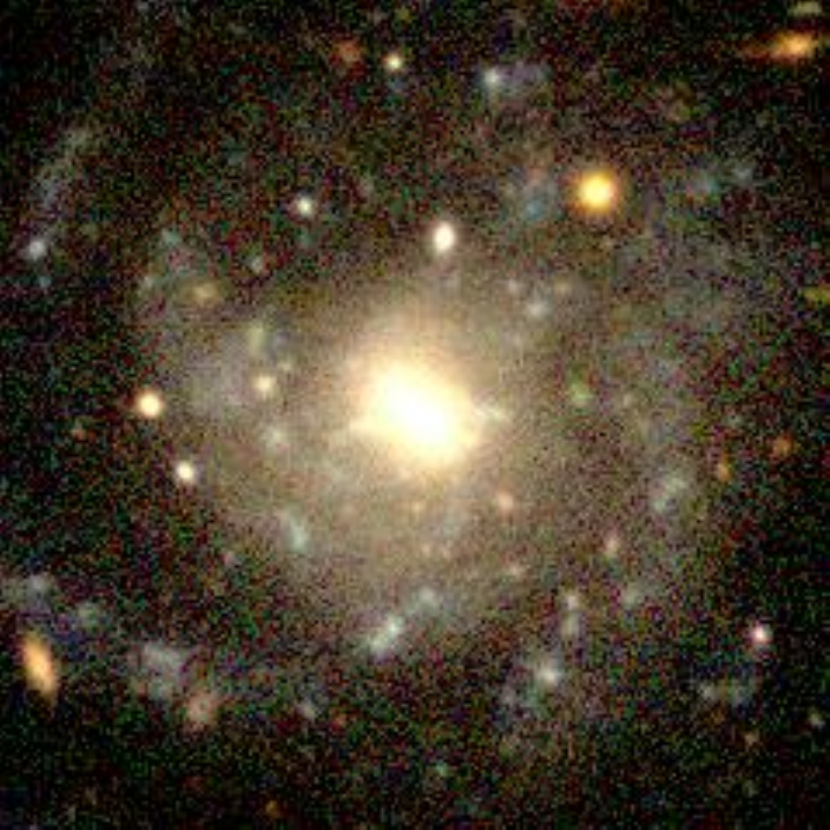}&
     \includegraphics[width=0.22\linewidth]{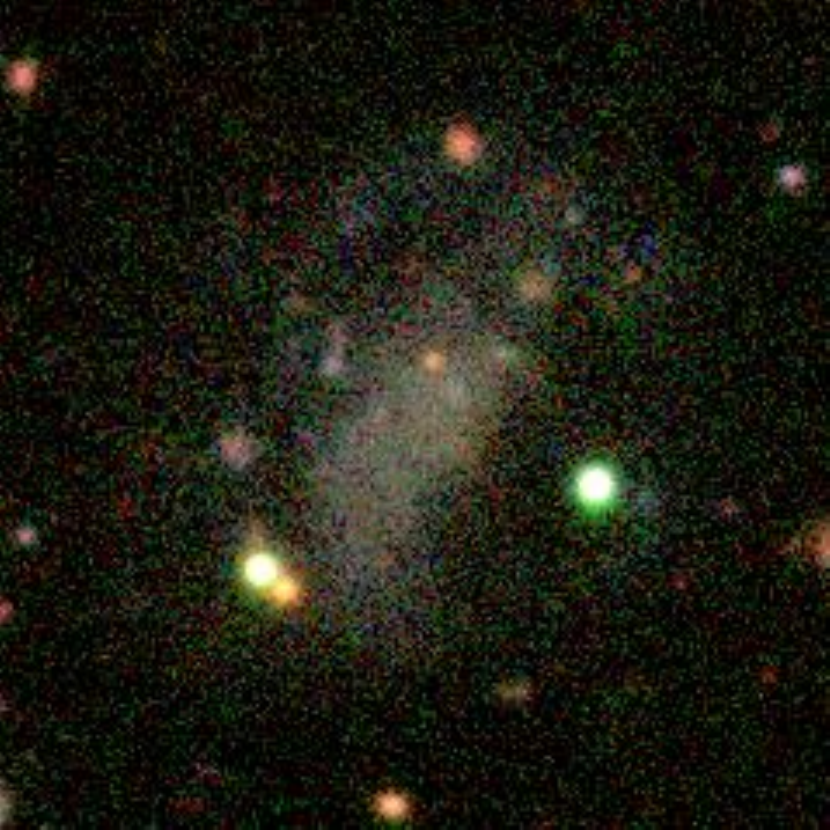}&
     \includegraphics[width=0.22\linewidth]{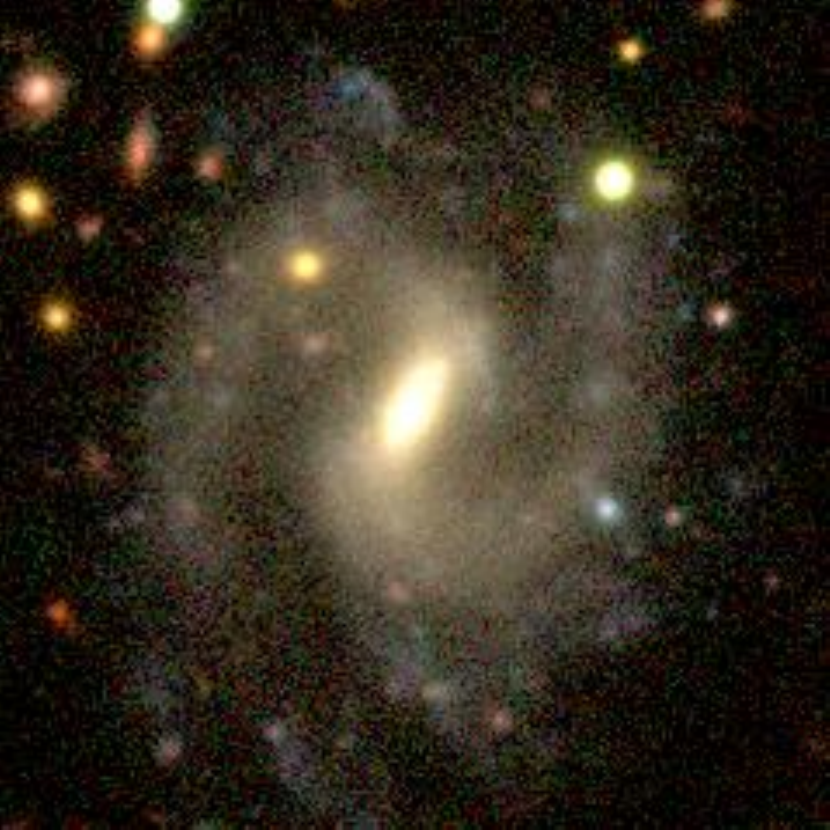}\\
\end{tabular}
\end{center}
\caption{Eight LSBGs of our sample in HSC wide field where the rgb color composite is based  on $g,r$ and $i$ band fluxes from HSC wide photometry. Image scale is 40 arcsec on a side.
\label{fig:LSBGface}}
\end{figure*}

HSC is a wide field imaging camera installed on the prime focus of the Subaru telescope \cite{2018PASJ...70S...1M, 2018PASJ...70S...2K}.
With this imaging camera and superb site conditions, seeing $\sim 0''.6$, HSC survey enables us to measure the shape and photometry for large number of galaxies out to redshift $z\sim 1.5$.
The HSC survey has three survey layers of Wide, Deep and Ultra deep, and in the Wide layer, 
we observe in five visible broad bands, $g, r, i, z$ and $y$.
The observation data set of the Wide layer for the first internal data release \textit{S16A} \cite{HSCDR1:2017} has survey areas of $\sim 200~{\rm deg^2}$ of the sky, which 
is eventually going to
reach to $1400~{\rm deg^2}$ in the final data release.
Limiting magnitudes for $5\sigma$ point-source detection of $g,r,i,z,$ and $y$-band are 26.5, 26.1, 25.9, 25.1, and 24.4, respectively.
The observation data is processed by an open-source software called \texttt{hscpipe}~(see \cite{Bosch+:2018} for the detail), which has been developed as a pipeline for the LSST data \citep{Axelrod+:2010, Juric+:2017}.

In \cite{2018ApJ...857..104G}, the LSBG catalog has been produced using the observation data set of the Wide layer with S16A data.
In order to avoid the limitation of survey areas by requiring all bands, they have used only three bands ($g,r,i$-band), because the survey progress is different by photometric bands.
Using \texttt{hscpipe}, they produce sky-subtracted co-add images and 
divide these images into equivalent rectangular regions called tracts (each tract has an area of $1.7~{\rm deg^2}$).
Further, tracts are divided into $9\times 9$ grids of patches and 
each patch is pixelized into $4200 \times 4200$ pixels.
To identify LSBG objects, the validity of the sky-subtraction is important
because of mean surface brightness of target objects fainter than the brightness of the night sky.
As described in \cite{Bosch+:2018}, an algorithm for the sky-subtraction used in \texttt{hscpipe} causes over-subtracting the background around extended sources~($>1'$).
Therefore, LSBGs found in \citep{2018ApJ...857..104G} can be biased against detection of LSBGs around such extended sources.

\texttt{hscpipe} is optimized to identify faint and small objects like distant galaxies.
If we use the pipeline for extended objects, the single extended object are decomposed into multiple child objects, so called ``shredding''.
Typically, LSBG in the HSC LSBG catalog is divided into more than 10 child objects.
Moreover, LSBG brightness is 
comparable to the sky background noise and thus the shape and surface brightness of LSBGs are likely misestimated.
Therefore in procedures to identify LSBGs, 
they have developed a pipeline based on the LSST codebase mainly,
instead of use of \texttt{hscpipe}.
Also they have used SExtractor \cite{1996A&AS..117..393B} to compose an initial catalog of LSBG candidates and then select them based on size and color measurements.
Finally \texttt{imfit} \cite{2015ApJ...799..226E} is used to refine their parameters.
Above procedures are carried out on a patch-by-patch basis. LSBGs are finally defined such that the mean surface brightness is larger than 24.3 mag~arcmin$^{-2}$. Finally, 781 LSBGs are detected within a HSC wide S16A footprint. The LSBGs identification procedure is summarized as follows, 

\begin{enumerate}
\item{} First of all, bright sources and associated diffuse lights are subtracted from images because they might mimic the LSBGs. Those subtracted region are replaced to the randomized background noise.
\item{} After smoothing with the Gaussian kernel (FWHM = $1''$), sources with $2.5'' < r_{1/2} < 20''$, where $r_{1/2}$ is the half-light radius, are extracted. 
Furthermore, by applying reasonable color cuts, optical artifacts and high-redshift galaxies are removed. 
\item{} By modeling the surface brightness profiles of LSBG candidates, astronomical false positive are removed. Finally, remaining false candidates, which is typically point-like sources with diffuse background lights, are removed by visual inspection. Final LSBG samples are 781 objects. 
\end{enumerate}
Basically, these procedures are applied to the $i$-bands images; however, in order to reduce the false detection including any artificial effects, all LSBG candidates are required to be detected in $g$-band images as well.
In this work, 
we use only eight LSBGs out of the 781 HSC LSBGs, where accurate distances to the LSBGs are measured.
\cite{2018ApJ...857..104G, 2018ApJ...866..112G}.
In Table \ref{tab:lsbg_param}, parameters of these eight LSBGs are summarized and in Figure \ref{fig:LSBGface}, each LSBG image used in this paper is shown.

\begin{table}
  \begin{center}
    \begin{tabular}{llllllll} \hline\hline
    LSBG ID & ($l$ [deg], $b$ [deg]) & redshift & $m_{i}$ & $m_{i, \rm err}$ & $g-r$ & $g-i$ & Distance[Mpc] \\ \hline
    171 & (62.628, -45.915) & 0.0439 & 17.55 & 0.24 & 0.49 & 0.77 & 186 \\
    285 & (178.250, -57.202) & 0.00581 & 17.47 & 0.24 & 0.44 & 0.6 & 24.9 \\
    456 & (351.210, 54.493) & 0.0286 & 17.18 & 0.24 & 0.4 & 0.57 & 122 \\
    464 & (348.724, 55.429) & 0.0257 & 16.95 & 0.24 & 0.37 & 0.58 & 110 \\
    575 & (224.099, 24.123) & 0.00695 & 18.45 & 0.24 & 0.29 & 0.3 & 29.7 \\
    613 & (339.731, 57.465) & 0.0244 & 19.17 & 0.24 & 0.19 & 0.29 & 104 \\
    729 & (336.533, 56.860) & 0.0251 & 17.57 & 0.24 & 0.36 & 0.55 & 107 \\
    750 & (276.818, 59.451) & 0.00862 & 18.38 & 0.24 & 0.24 & 0.33 & 36.9 \\ \hline \hline
    \end{tabular}
  \end{center}
  \caption{
  Parameters of LSBGs in our analysis. Each parameter represents object ID, Galactic longitude, Galactic latitude, redshifts, $i$-band magnitudes, errors of $i$-band magnitudes, color diagram for $g-r$, $g-i$ and comoving distances of LSBGs, respectively.
    \label{tab:lsbg_param}}
\end{table}

\subsection{LAT data reduction}
\label{ssec:lat}
In this work, we use 8 years~(2008-08-04 to 2016-08-02) 
of \emph{Fermi}-LAT observations. We have selected the photon event class 
{\tt P8R3 SOURCE}, which is the recommended class of events for point source analysis. Only events within the energy range 500 MeV to 500 GeV are selected, and these are further binned in 26 logarithmically spaced energy bins in our analysis. The LAT possesses a low angular resolution for energies $\lesssim 500$ MeV which could introduce a bias in the analysis due to possible point source confusion, while at high energies the photon count statistics decreases. The chosen low energy limit is a compromise between sensitivity and
statistics. For each LSBG shown in Tab.~\ref{tab:lsbg_param} we select a patch of the sky of size $10^{\circ}\times 10^{\circ}$ 
with centroid in each target object.
In the construction of the counts maps we have employed spatial bins of size $0.1^{\circ}$ and the corresponding instrument response functions~(IRFs) for our photon class event~ \cite{2012ApJS..203....4A}, 
{\tt P8R3\_SOURCE\_V2}
\footnote{\url{https://fermi.gsfc.nasa.gov/ssc/data/analysis/documentation/Cicerone/Cicerone_LAT_IRFs/IRF_overview.html}}. As recommended in the \texttt{cicerone}\footnote{\url{https://fermi.gsfc.nasa.gov/ssc/data/analysis/documentation/Cicerone}}, we applied the quality cuts {\tt DATA\_QUAL>0 \&\& LAT\_CONFIG==1}.
Furthermore, we reject events with zenith angles larger than $100^{\circ}$ to avoid contamination of photons produced by interaction of cosmic rays with the Earth's atmosphere. The analysis of the LAT data was done with \texttt{fermipy}\footnote{\url{https://fermipy.readthedocs.io/en/latest/}} \cite{2017ICRC...35..824W}, which is an open-source software package based on the {\tt Fermi Science Tools}\footnote{\url{https://fermi.gsfc.nasa.gov/ssc/data/analysis/software/}} (\texttt{v11r5p3}), including some additional high-level tools for common analysis tasks. 


In our analysis, we first perform maximum likelihood runs to estimate the parameters for all $\gamma$-ray sources in our ROIs. 
The Galactic foreground and background $\gamma$-ray emission is modeled with the standard diffuse Galactic foreground model,
gll\_iem\_v07.fits, and the isotropic emission model,
iso\_P8R3\_SOURCE\_V2\_v01.txt,
respectively. 
As for the resolved $\gamma$-ray sources, we derive them from the 4FGL catalog \cite{4FGL}.
The amplitude and spectral shape of all the $\gamma$-ray sources lying in our ROIs were varied in our maximum likelihood runs while the sources which reside outside our ROIs and within region of $15^\circ \times 15^\circ$ centered on LSBG were fixed to the their catalog values.
Figure~\ref{fig:ph_count} displays the photon count maps around each LSBG as observed by the \emph{Fermi}-LAT. The corresponding residual maps obtained after subtracting the best-fit background/foreground emission model are also shown in the same figure. These images are constructed using the full energy range (500 MeV to 500 GeV). 
We note that photons counts above a few of tens of GeVs hardly contribute to the overall observed emission.

Following the procedure of~\cite{2015PhRvL.115w1301A}, we search for excess emission at the position of each LSBG. This is done separately in each energy bin in order to derive flux constraints that are not sensitive to their assumed spectrum. To obtain convergence in our maximum likelihood fits, we utilize the following parameter relaxation  method: First, we optimize the amplitudes of all $\gamma$-ray sources in order of higher intensity sources down to fainter ones.
Next, we repeat our maximum likelihood run starting from the updated source parameters and freeing all the spectral shape parameters. The last step of our procedure consists in running the \texttt{fermipy} pipeline for the flux upper limits calculations. 

In the computation of the flux upper limits we follow the prescription given in the 2FGL catalog~\cite{2FGL}. 
Namely, we employ a Bayesian~\cite{1983NIMPR.212..319H} method recommended for analyses of very faint $\gamma$-ray point sources. For a rigorous application of this method, we first checked that our LSBGs had test statistics~(TS) values of less than one in most energy bins.
In addition, since the point-source fluxes are assumed to be positive, the likelihood function can be approximated to a $\chi^2/2$ distribution. In this sense, the 95\% confidence level (C.L.) flux upper limits are obtained by varying the spectral normalization parameter of the putative LSBGs until $\log \mathcal{L}_0 - \log \mathcal{L} \sim2$ (where $\mathcal{L}_0$ is the likelihood for zero-flux). \DvI{(see Section 3.5 of \cite{2FGL})}.

\begin{figure}
 \begin{center}
 \hspace{-5mm}
   \includegraphics[width=15cm]{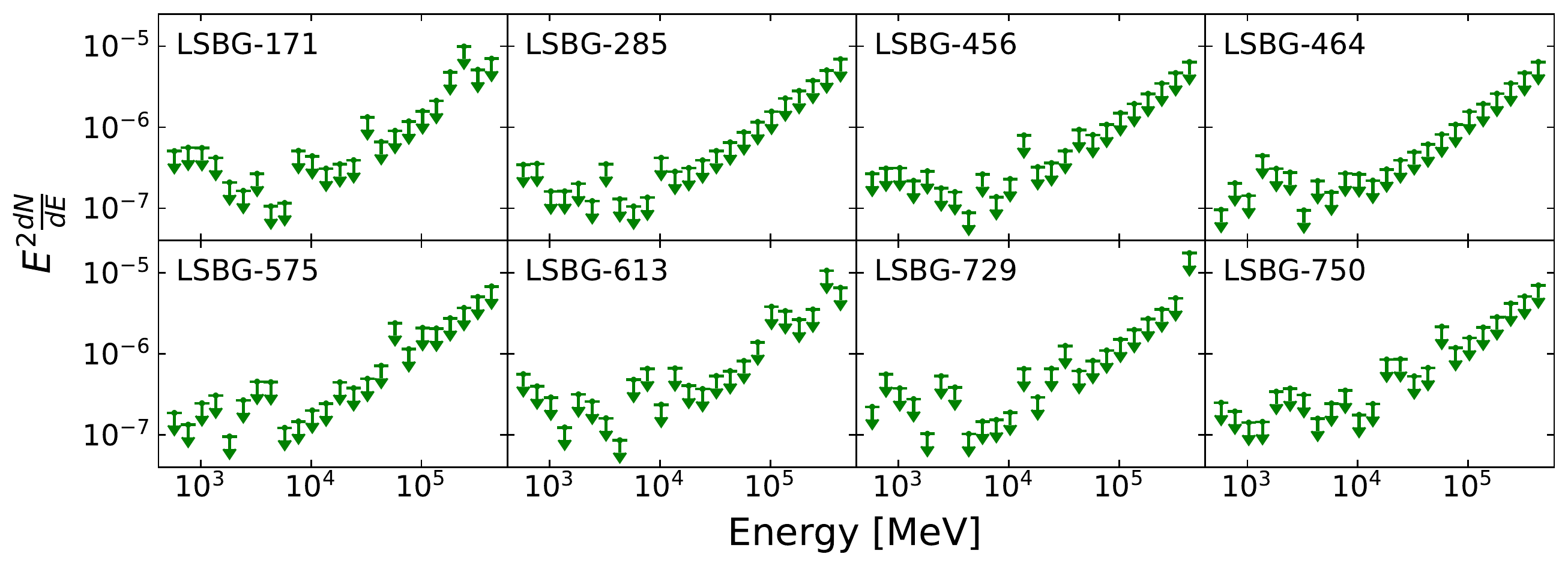}
 \end{center}
 \caption{
 Flux upper limits (at the $2\sigma$ level) for the sample of eight LSBGs used in our analysis.
 The limits are obtained using a bin-by-bin maximum likelihood analysis assuming that the LSBGs are $\gamma$-ray point sources described by a simple power-law spectrum with a fixed slope of $-2$ at each energy bin. }
 \label{fig:lsbg_likepro}
\end{figure}

Figure \ref{fig:lsbg_likepro} shows the likelihood profiles for all our target LSBGs. 
We note that some of our ROIs overlap with each other, however, in the computation of the combined limits with a joint likelihood procedure we assume that all of our ROIs are independent. 
Details on the methods utilized for converting flux upper limits into constraints for the DM annihilation cross-section are given in the next section.

\iffigure
\begin{figure}
 \begin{center}
   \begin{tabular}{cccc}
   \hspace{-0.7cm}
   \includegraphics[width=4.1cm]{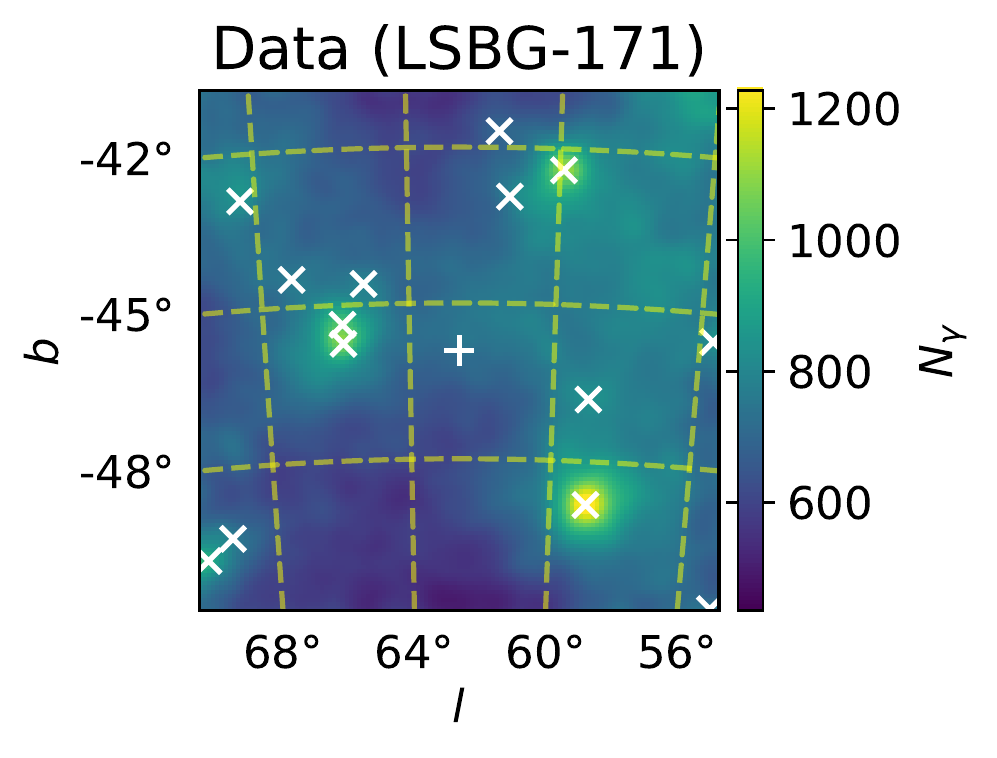}&
   \hspace{-0.6cm}
   \includegraphics[width=4.1cm]{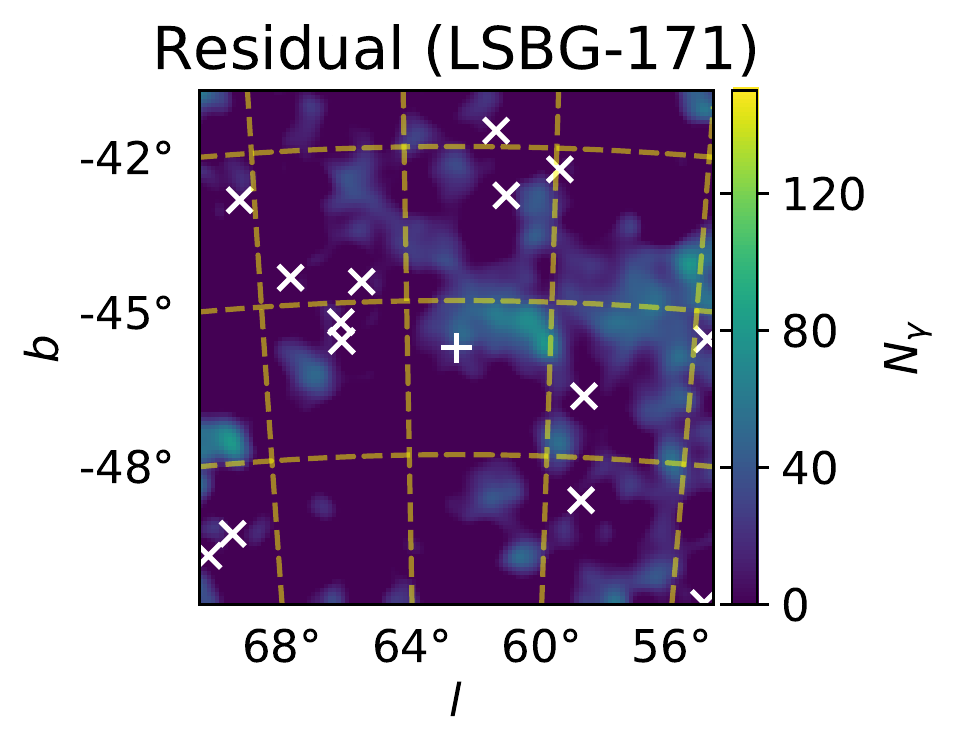}&
   \hspace{-0.6cm}
   \includegraphics[width=4.1cm]{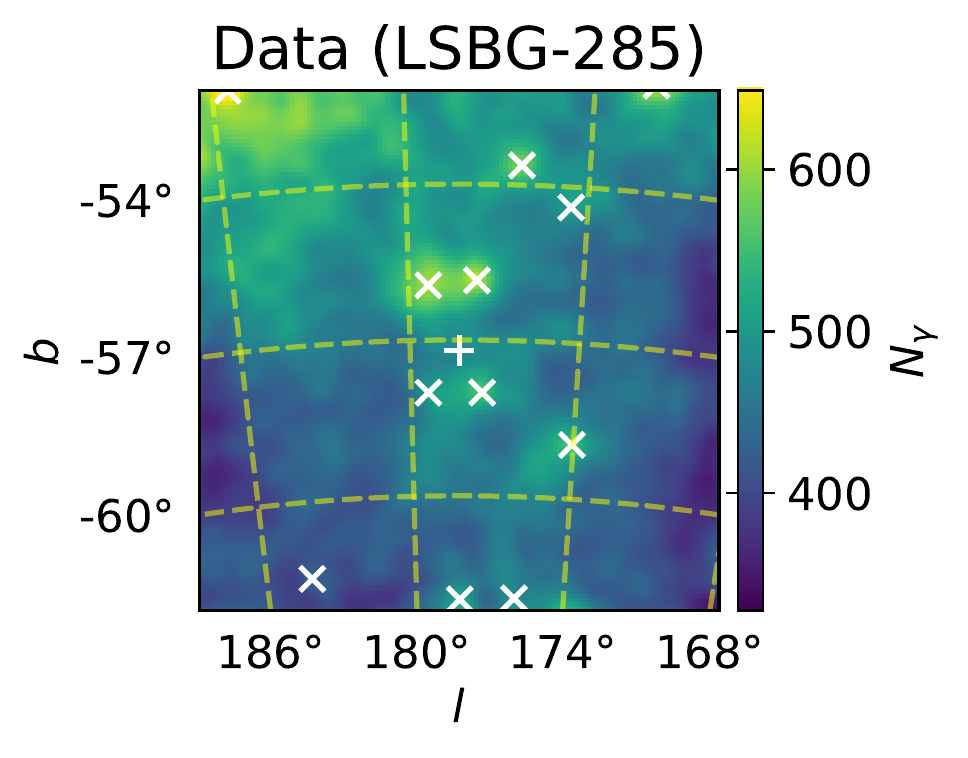}&
   \hspace{-0.6cm}
   \includegraphics[width=4.1cm]{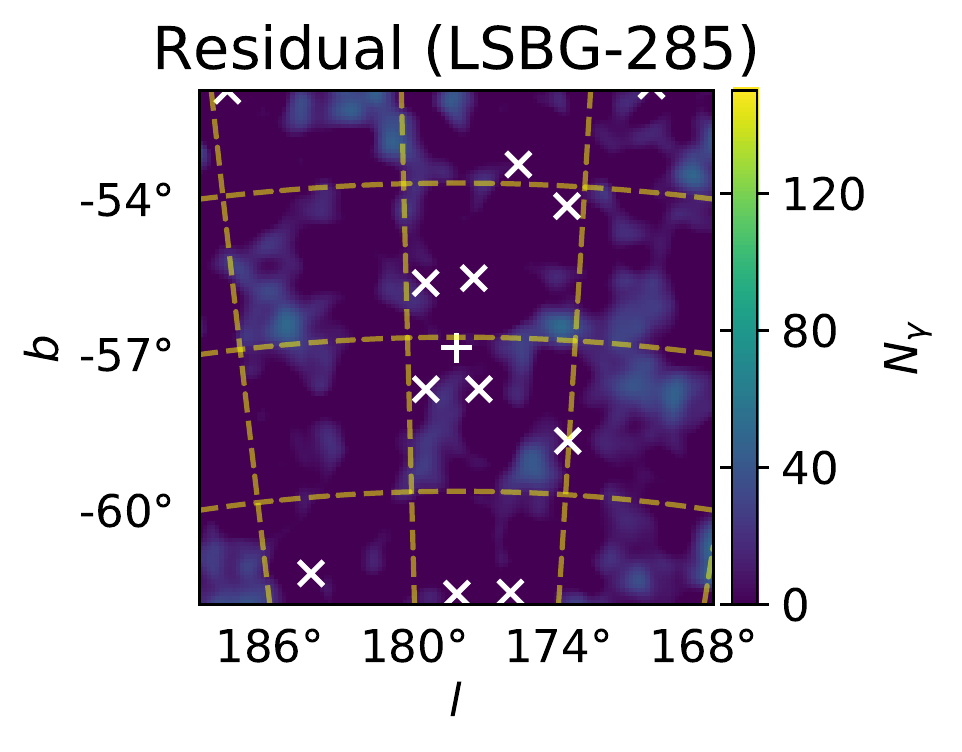}\\
   \hspace{-0.7cm}
   \includegraphics[width=4.1cm]{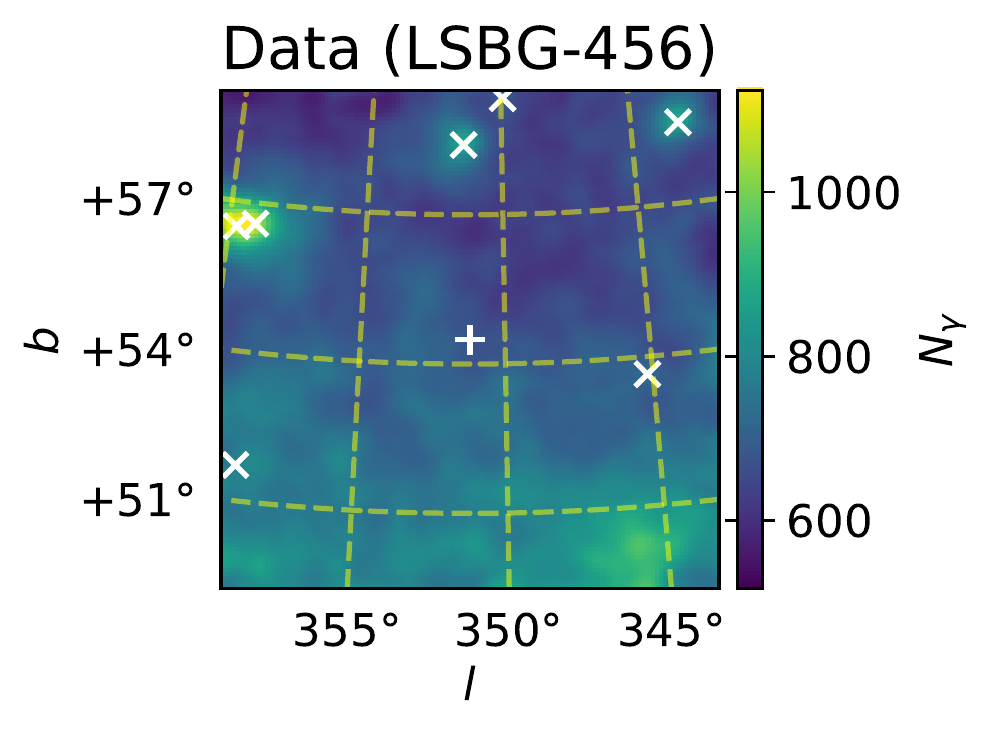}&
   \hspace{-0.6cm}
   \includegraphics[width=4.1cm]{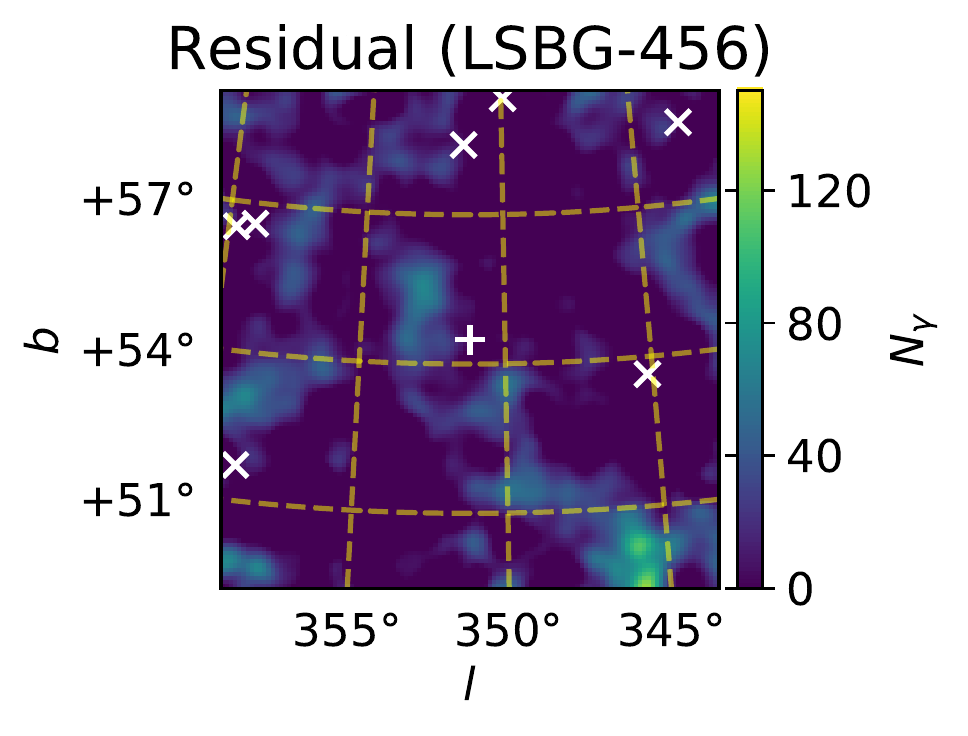}&
   \hspace{-0.6cm}
   \includegraphics[width=4.1cm]{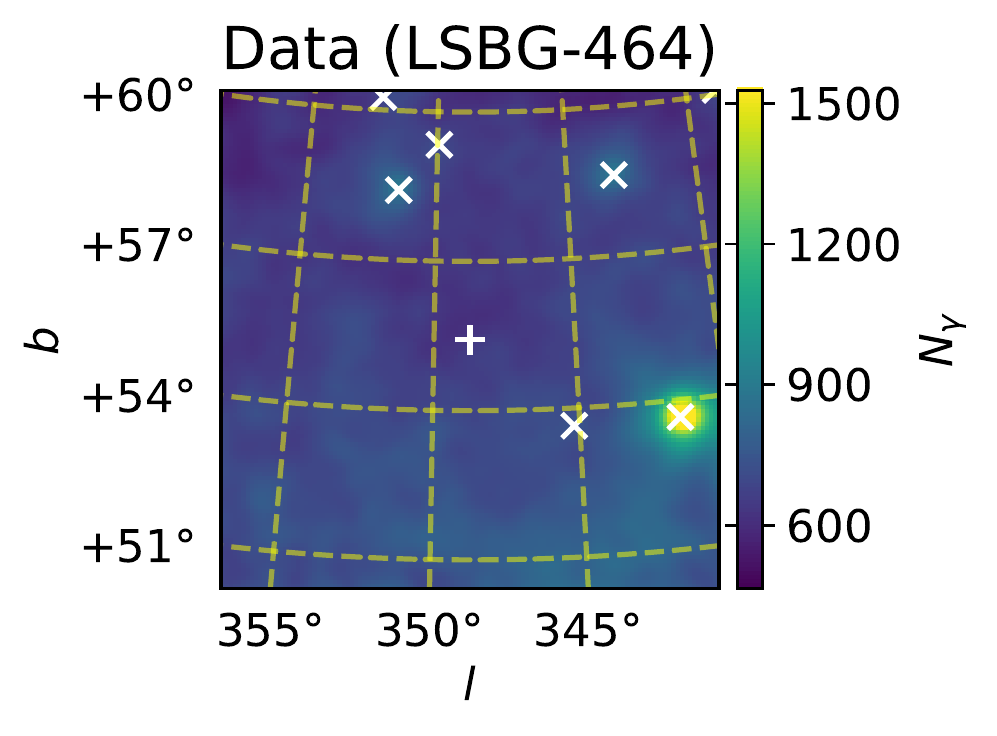}&
   \hspace{-0.6cm}
   \includegraphics[width=4.1cm]{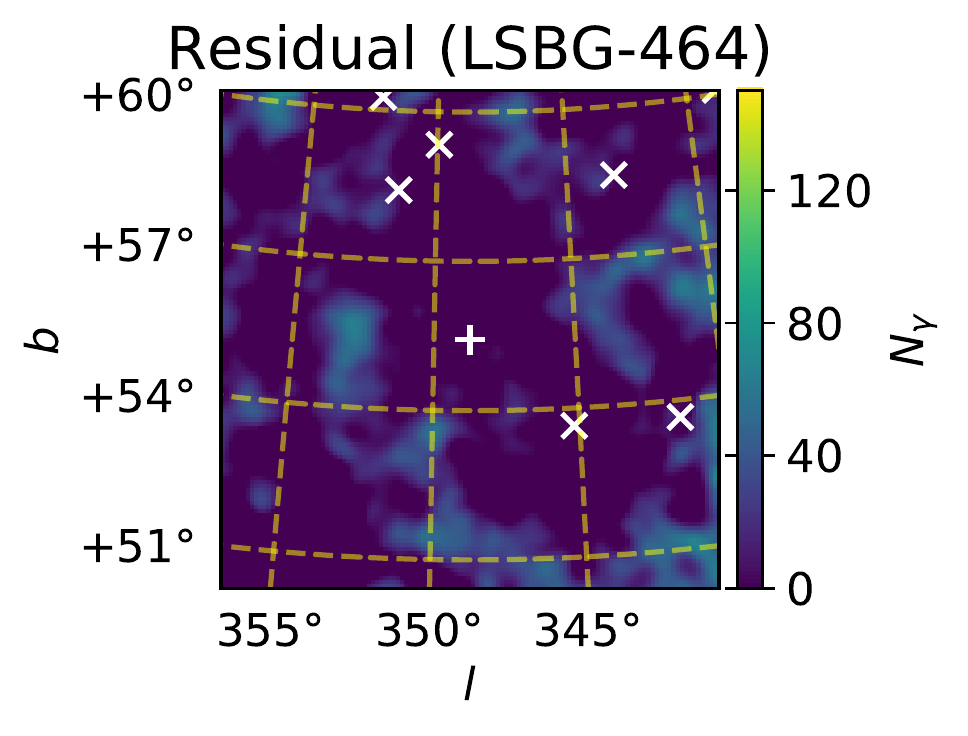}\\
   \hspace{-0.7cm}
   \includegraphics[width=4.1cm]{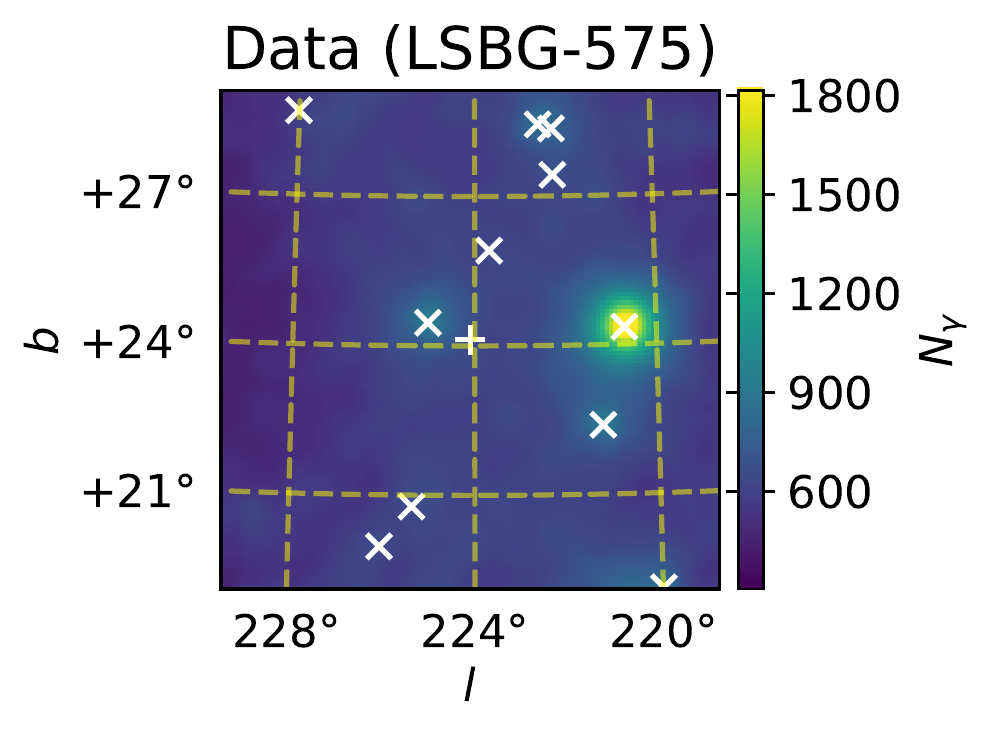}&
   \hspace{-0.6cm}
   \includegraphics[width=4.1cm]{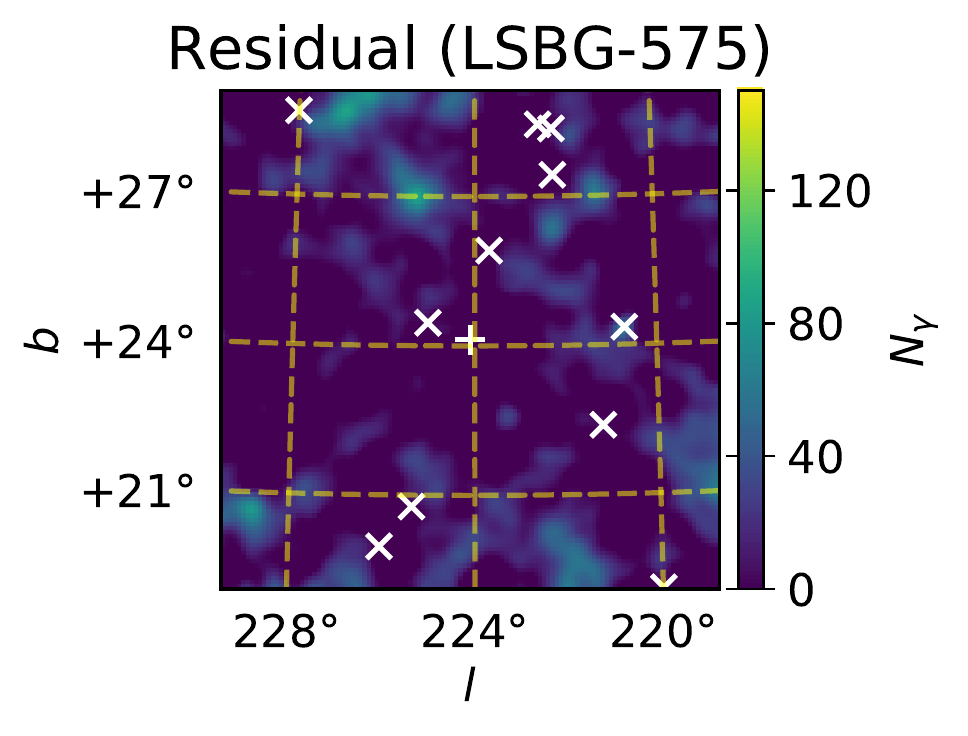}&
   \hspace{-0.6cm}
   \includegraphics[width=4.1cm]{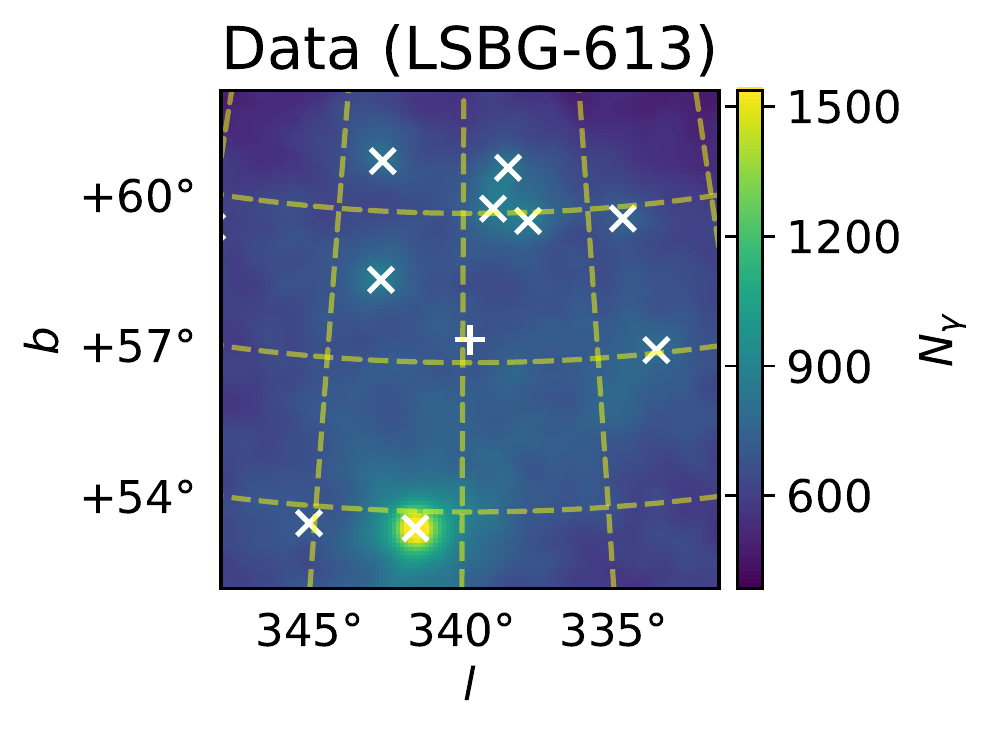}&
   \hspace{-0.6cm}
   \includegraphics[width=4.1cm]{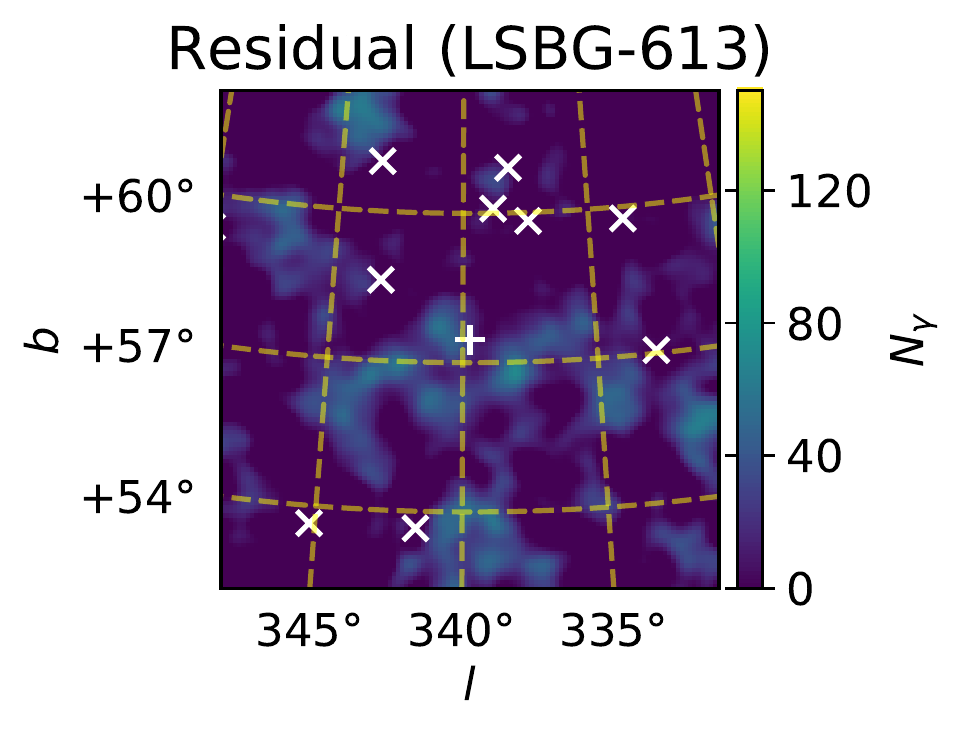}\\
   \hspace{-0.7cm}
   \includegraphics[width=4.1cm]{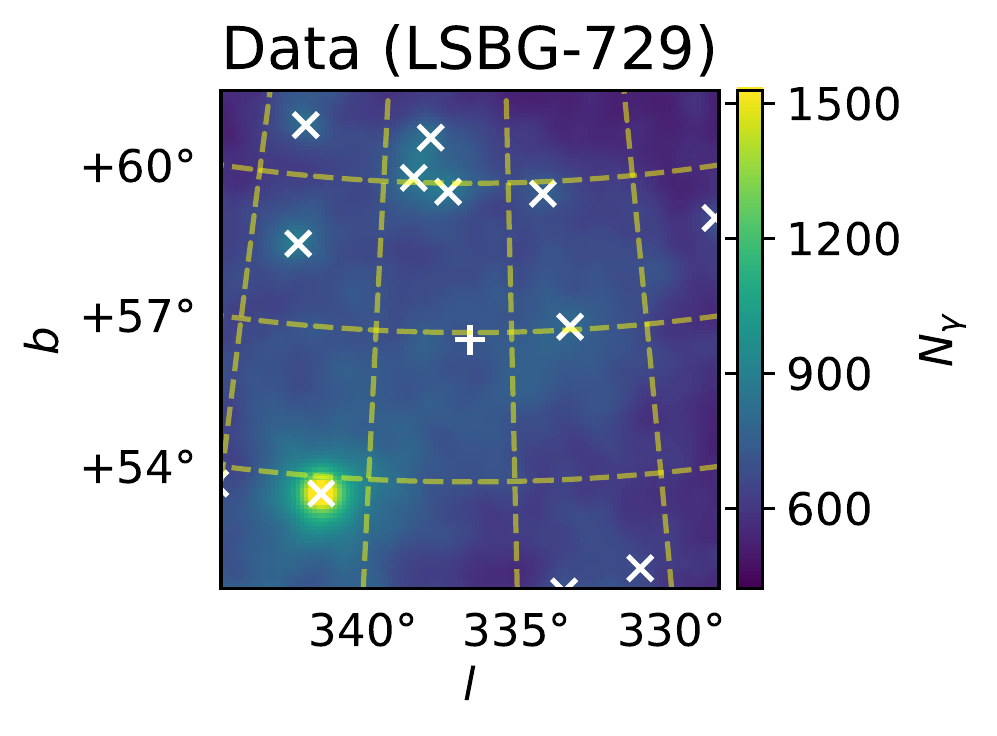}&
   \hspace{-0.6cm}
   \includegraphics[width=4.1cm]{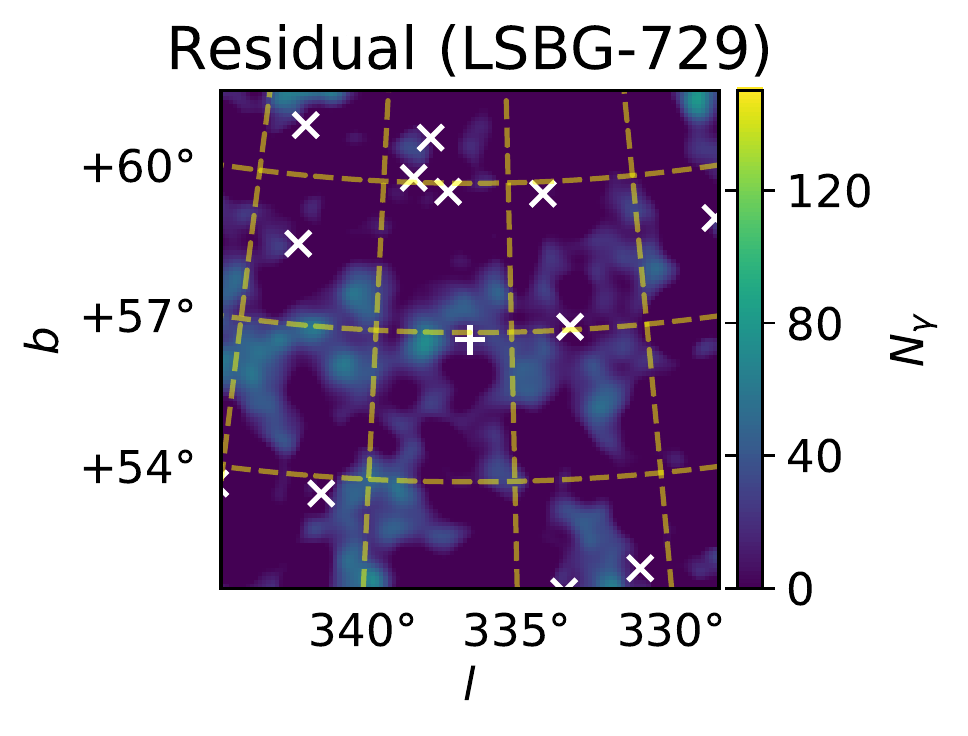}&
   \hspace{-0.6cm}
   \includegraphics[width=4.1cm]{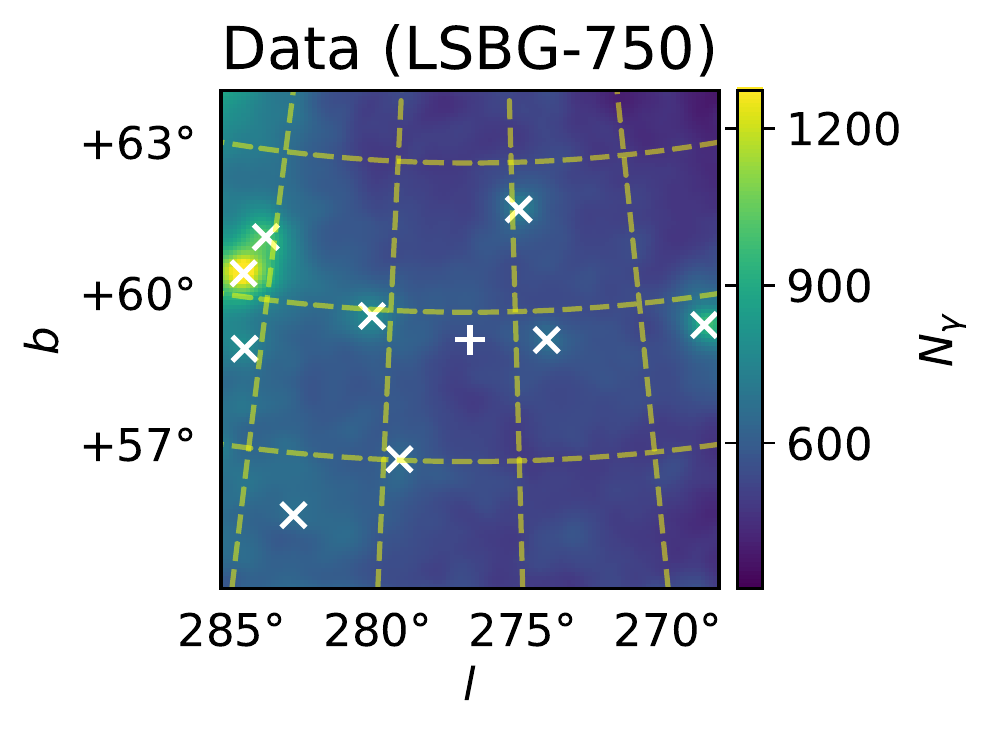}&
   \hspace{-0.6cm}
   \includegraphics[width=4.1cm]{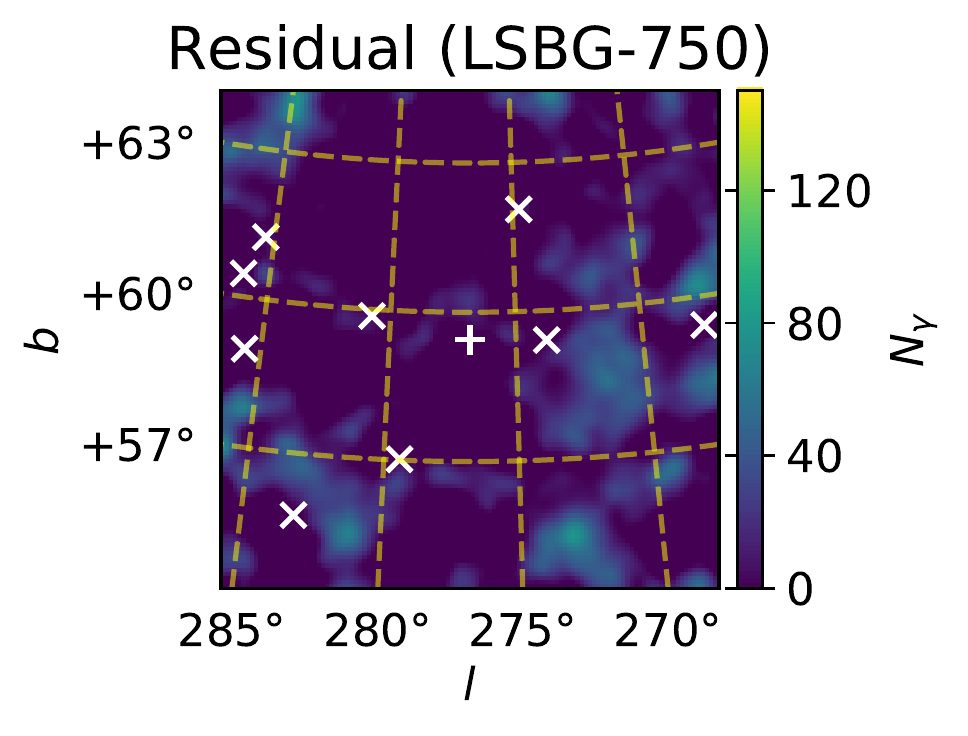}\\
  \end{tabular}
 \end{center}
 \caption{
  Photon and residual counts maps (for the energy range $[0.5,500]$ GeV) centered at the position of each LSBG considered in this work (see also  Tab.~\ref{tab:lsbg_param}). Each patch encompasses a region of $10^\circ \times 10^\circ$ of the sky and is displayed in Galactic coordinates. The position of each LSBG is represented by a cross ``+'' while 4FGL catalog point sources lying in our ROI are shown as an ``$\times$''.
  The color scale represents the number of observed (residual) photon counts. Residuals are obtained by subtracting each corresponding best-fit background/foreground model from the observed photon count maps.
 }
 \label{fig:ph_count}
\end{figure}
\fi

\section{Methods}
\label{sec:model}
\subsection{Modelling of the Dark Matter annihilation signal}
\label{ssec:model}
In this section, we provide details on our modelling methods for the expected $\gamma$-ray emission from DM annihilation as well as 
on the pipeline used to constrain the DM model parameters.

The predicted $\gamma$-ray annihilation flux from each LSBG can be written as
\begin{equation}
\frac{d\Phi_{\gamma}}{dE_{\gamma}} = J \times \frac{\langle \sigma v \rangle}{8\pi m_{\chi}^{2}}
\sum_{i}Br_{i}\frac{dN_{i}}{dE_{\gamma}'}\Bigg{|}_{E_{\gamma}'=(1+z)E_{\gamma}},
\label{eq:ann_flux}
\end{equation}
where $z$ is the redshift of the LSBGs, $\langle \sigma v \rangle$ is the ensemble average of the annihilation cross-section and relative velocity of DM particles, $E_{\gamma}, E_{\gamma}'$ are observed and emitted $\gamma$-ray energy, $m_{\chi}$ is DM mass, $dN/dE_{\gamma}$ is the $\gamma$-ray energy spectrum and $Br_{i}$ is the branching ratio in the $i$-th annihilation channel, respectively. The J-factor,
$J$ 
is fully specified by astronomical parameters of target halos, such as the halo mass or distance to the halo. We note that any astronomical factors to be considered for the $\gamma$-ray flux are absorbed in the J-factor.
Other factors in the right-hand side of Eq. (\ref{eq:ann_flux}) are fully given by the particle physics nature of the DM candidate.
In general, DM 
particles can self-annihilate in various final state particles such as the 
$b\bar{b}, \tau^+\tau^-, \mu^+\mu^-$ and $W^+W^-$ as well as $\gamma\gamma$ and $\gamma Z$, 
which in turn can produce secondary $\gamma$-ray emission through their subsequent cascade decay. 
In our analysis, we consider the single annihilation channel $b\bar{b}$, 
which is a representative channel and benchmark to assess 
the sensitivity of the LAT to a putative DM signal from our search targets. The DM spectra is obtained with the \texttt{DMFIT}~\cite{Jeltema:2008hf} package provided within \texttt{fermipy}. The \texttt{DMFIT}\footnote{\url{https://fermi.gsfc.nasa.gov/ssc/data/analysis/scitools/gammamc_dif.dat}} itself provides an interpolation to DM $\gamma$-ray spectra tables extracted from {\tt DarkSUSY}~\cite{2018JCAP...07..033B}.

Amongst the various effects expected to impact the $J$-factors, the mass estimate should include the largest uncertainty. Here we describe the way to estimate the halo mass for the eight LSBGs with HSC optical data. 
In particular, we convert the observed $g, r$ and $i$ band magnitudes into $V$ band magnitude using 
\citep{2005AJ....130..873J},
\begin{equation}
    V = g - 0.59(g-r) -0.01.
\end{equation}
%
Given the luminosity distance $d$, we can convert apparent magnitude to absolute magnitude $M_V$, by using the well known relation $M_V = V + 5 - 5\log_{10}d$. Also, by assuming the mass-to-light ratio to be unity \citep{2008MNRAS.390.1453W} we obtain
%
%
\begin{equation}
     M_* = \frac{M_{\odot}}{L_{\odot}}L_{\rm LSBG}.
\end{equation}
Finally, we apply the stellar to halo mass ratio \cite{2013MNRAS.428.3121M} for the crude estimate of the total halo mass $M_{\rm halo}$ of our LSBG sample.
%
%
In Table \ref{tab:Jfact}, stellar mass and halo mass of the eight LSBGs are summarized. Below we describe how the $M_{\rm halo}$ estimates are implemented in the J-factor calculations.

The J-factor is defined as the line of sight integral of the squared DM density profile,

\begin{equation}
J = [1 + b_{\rm sh}(M_{\rm halo})] \int_{s} ds' \int_{\Omega} d\Omega' \rho^{2}_{\rm DM}(s',\Omega')
\label{eq:j_factor}
\end{equation}
where $\Omega$, $\rho_{\rm DM}$ and $M_{\rm halo}$ are the solid angle of the target object, the DM density profile and the halo mass, respectively.
The the boost factor $b_{\rm sh}(M_{\rm halo})$ takes into account the excess of the annihilation rate due to the substructure present in the halo.
In our analysis, we employ the model proposed in Ref.~\cite{Hiroshima+2018} for the calculation of the boost factor and assume $b_{\rm sh}(M_{\rm halo}) = 1$ for all 8 LSBGs.
As for the dark matter profile, we assume a smooth NFW density profile \cite{NFW},
\begin{equation}
\rho_{\rm DM}(r) = \frac{\rho_{s}}{c r/r_{\rm vir}\left[(c r/r_{\rm vir}) + 1\right]^{2}},
\end{equation}
where $r_{\rm vir}$ is the virial radius, $\rho_s$ and $c$ are the normalization and concentration. We adopt the mass-concentration relation obtained in \cite{2014MNRAS.441.3359D}. The parameters $\rho_s$ and $c$ can be uniquely determined once the DM halo mass is specified.
%
%
It can be shown that the volume integral of the $\rho^2_{\rm DM}$ 
can be reduced to
\begin{equation}
\int_{V} dV \rho_{\rm DM}^{2}(r) = \frac{M_{\rm halo}}{9} \Delta \rho_c(z)c^{3}\left( \log(1+c) - \frac{c}{1+c} \right)^{-2}\left( 1 - \frac{1}{(1+c)^{3}} \right),
\label{eq:rho2}
\end{equation}
where $\Delta$ is the overdensity of spherical collapse to be assumed as $200$ and and $\rho_c$ is the critical density of the Universe.
Since the angular size the LSBGs considered in this study is much smaller than the PSF of the LAT instrument in all energy bands, we can in practice treat all the LSBGs as $\gamma$-ray point sources in our analysis pipeline.
It follows that the line of sight integral can be replaced as,
\begin{equation}
    \int ds \int d\Omega \rho^2_{\rm DM}(s,\Omega)
    \rightarrow
    \int dV \rho^2_{\rm DM}(r) / d_A^2.
    \label{eq:rho2_int}
\end{equation}  
Thus, eq.(\ref{eq:j_factor}) can be simplified as
\begin{align}
 J 
 &= [1 + b_{\rm sh}(M_{\rm halo})] \frac{M_{\rm halo}}{9} \Delta \rho_{c,z}c^{3}\left( \log(1+c) - \frac{c}{1+c} \right)^{-2}\left( 1 - \frac{1}{(1+c)^{3}} \right) / d_{\rm A}^{2},
 \label{eq:finalJ}
\end{align}
where $d_A$ is the comoving angular diameter distance to the LSBG. 
To evaluate the uncertainty on the $J$-factor, we perform Monte-Carlo error estimation by assuming 1-$\sigma$ Gaussian error, $\Delta \log_{10}c=0.1$ \citep{2013MNRAS.428.3121M}. 
Halo mass uncertainties can be converted from $i$-band magnitude uncertainty from Table \ref{tab:lsbg_param} which can read as 1-$\sigma$ Gaussian errors with $\Delta \log_{10} M_{\rm halo} \sim 0.4$.
We ignore the error on the distance as it is sufficiently accurately measured.
\OMvI{Furthermore, astrophysical uncertainties in the matter profile of LSBGs result in J-factor uncertainties of at most 0.2 dex \cite{2018PhRvL.120j1101L}.}

\begin{table}
  \begin{center}
    \begin{tabular}{cccc} \hline\hline
    LSBG ID & ${\rm log}_{10}(M_{*}[M_{\odot}])$ & ${\rm log}_{10}(M_{\rm halo}$$[M_{\odot}]$) & ${\rm log}_{10}(J [{\rm GeV}^{2}{\rm cm}^{-5}]$) \\ \hline
    171 & 9.3 & 11.1 & 14.4 \\
    285 & 7.6 & 10.4 & 15.7 \\
    456 & 9.1 & 11.1 & 14.7 \\
    464 & 9.1 & 11.0 & 14.8 \\
    575 & 7.4 & 10.3 & 15.5 \\
    613 & 9.3 & 10.7 & 14.7 \\
    729 & 9.3 & 10.7 & 14.8 \\
    750 & 7.6 & 10.4 & 15.4 \\ \hline \hline
    \end{tabular}
  \end{center}
  \caption{
  \label{tab:Jfact}
  Stellar mass, halo mass and $J$-factor value for eight LSBGs.
  Each parameter is a median value in the Monte Carlo simulation.
  Errors of stellar mass, halo mass and $J$-factor are $\sim0.2, \sim0.4$ and $\sim0.7$ in logarithmic, respectively.
  }
\end{table}

\subsection{Composite Analysis}
\label{ssec:comp_ana}
As it can be inferred from Eq.~\ref{eq:ann_flux}, the predicted DM energy annihilation spectra is independent of the search target, only the J-factors present differences in our sample of LSBGs. This characteristic allows us to combine data from all individual LSBGs in order to set stronger constraints on the DM model parameters. After first analyzing every individual LSBG, a composite analysis is subsequently performed following the methods employed by the \textit{Fermi} team in dSphs~\cite{2015PhRvL.115w1301A}. Specifically, our pipeline adds together the photon counts from the signal and background regions for each LSBG and then computes upper limits on the signal region following the same prescription as for individual sources. We note that this composite likelihood technique takes into account that every LSBG has a different J-factor.
All the LSBGs are found to have very low statistical significance detection in the \textit{Fermi} data. 
Therefore, in the computation of the 95\% C.L. upper limits of the DM annihilation cross-section, we employ the Bayesian~\cite{1983NIMPR.212..319H} procedure recommended in the 2FGL catalog~\cite{2FGL} for analyses of dim point sources. 

Using our bin-by-bin method we perform likelihood scans as a function of putative DM fluxes associated to each LSBG in every individual energy bin. Since all the likelihood values obtained at each ROI are assumed to be independent of each other, the joint likelihood ${\mathcal L}_{\rm stack}$ for the full sample of targets can be written as 
\begin{equation}
 \log {\mathcal L}_{\rm stack}({\cal D}|\{ \sv, J_1, J_2,...\}) = \sum_{i,j} \log {\mathcal L}^{\rm ann}_{i,j, }({\cal D}_j|\{ \sv, J_j\}),
 \label{eq:stackL}
\end{equation}
where ${\cal L}^{\rm ann}_{i,j}$ is the log-likelihood value obtained for the $j$-th LSBG, $i$-th energy bin and $J_j$ is the $J$-factor of one particular LSBG. 
In this formulation, we assume that our all LSBGs are statistically independent from each other. Since our sample of LSBGs are treated as point sources, this assumption is correct if the angular separations between LSBGs are larger than the LAT PSF at the energy range in question. We will further discuss in Sec.~\ref{ssec:DM_const}.


As described in Section \ref{ssec:lat}, in our procedure we assume that the flux of the LSBGs is positive definite. This implies that in the limit of large number counts, the data is well described by a $\chi^2/2$ distribution. In this sense the 95\% C.L. upper limits on the velocity averaged DM cross section can be obtained when the total likelihood  $\Delta\log {\mathcal L}_{\rm stack}({\cal D}|\sv) \sim 3.8/2$ (for 1 degree of freedom). We estimate the astrophysical uncertainties associated  to the upper limits computed this way by performing Monte Carlo simulations of the J-factor distributions for the eight LSBGs considered in this work.

\section{Result}
\label{sec:result}
\subsection{Dark Matter Constraints }
\label{ssec:DM_const}
\begin{figure}
 \begin{center}
   \includegraphics[width=12cm]{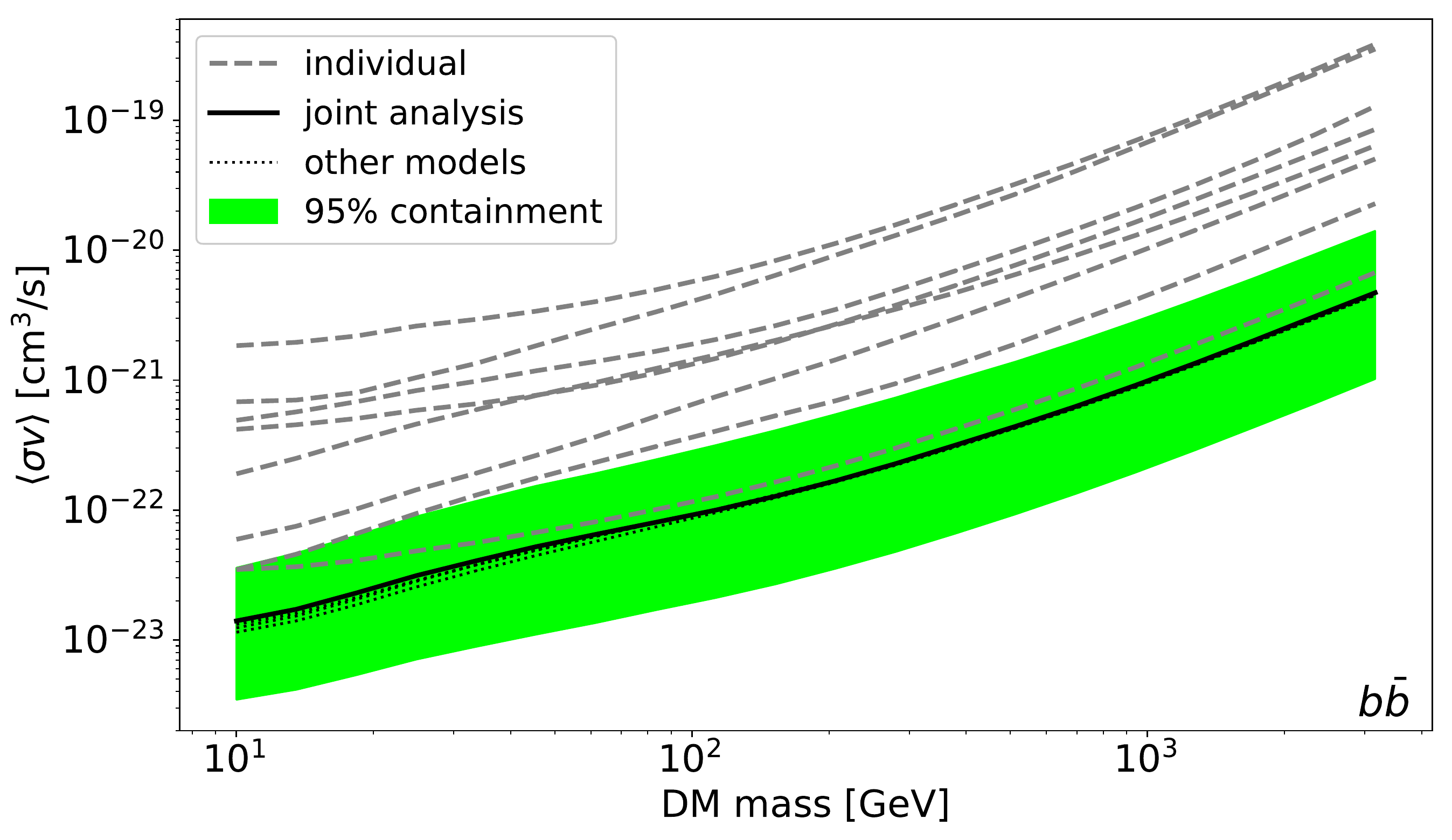}
 \end{center}
 \caption{
\OMvI{95\% C.L. upper limits on the DM annihilation cross-section in the $b\bar{b}$ channel. The black
solid line represents the upper limit obtained with a joint analysis assuming the median J-factor values of each LSBG, while the grey dashed lines display the upper limit obtained for every individual LSBG in our 8 objects sample. The black dotted lines show the joint upper limits obtained after replacing our baseline galactic diffuse emission model with 3 alternative background models simulated with GALPROP. The green band shows the uncertainty of the upper limit due to uncertainties in the J-factor. This is estimated by Monte Carlo sampling the individual J-factors. For more details on the Monte Carlo simulation see the text
in Section~\ref{ssec:comp_ana}.} }
 \label{fig:sigmav}
\end{figure}

As shown in Figure~\ref{fig:lsbg_likepro}, the sample of LSBGs considered in this work were not significantly detected in $\gamma$-rays. Therefore, we use them to impose constraints on the DM annihilation cross section. The computation of the upper limits is done using the methods explained in the previous section. Figure~\ref{fig:sigmav} displays the 95\% C.L. upper limits on $\sv$ as obtained for each individual LSBG (dashed lines) using their respective median J-factor. The joint upper limit for the full sample (black solid line) was obtained with the joint likelihood method described in Eq.~\ref{eq:stackL}.
The green band displays the impact on our limits due to astrophysical uncertainties in the DM model parameters. As discussed earlier, these come mainly from uncertainties in the halo mass and the matter concentration of the LSBGs. As it can be seen, the objects LSBG-285 and LSBG-575 place the strongest constraints on $\sv$ and also provide dominant contributions to the joint constraint.
Even after stacking over the full sample of eight LSBGs, our joint constraints are weaker than the ones obtained using more traditional targets like dSphs or nearby galaxy groups and clusters of galaxies \cite{2015PhRvL.115w1301A, 2018PhRvL.120j1101L}.
We note that in our stacking procedure we assume that the log-likelihood values obtained for each LSBG are  statistically independent. However we found that there are two pairs of LSBGs -- (LSBG-456, LSBG-464) and (LSBG-613, LSBG-729) -- that are separated by only $\sim 2^{\circ}$. This separation is comparable to the size of LAT PSF in our low energy bins and it is possible that these objects might bias our results. In order to estimate the impact that this assumption had on our results we computed the corresponding cross-covariance for both pairs of LSBGs. In particular, we found that the impact on our upper limits was at the $\sim 10\%$ level. This demonstrates that the statistical independence assumption is appropriate for our analysis.

In addition, we have estimated the systematic uncertainties introduced by our Galactic diffuse emission model. We followed the rigorous approach recommended in Ref.~\cite{2015ApJ...799...86A}. In that reference three different Galactic diffuse emission model were constructed using the Cosmic Rays (CR) propagation code~\texttt{GALPROP}\footnote{\url{http://galprop.stanford.edu}}. The three different model named Model A, B and C encapsulate a wide range of uncertainties in the interstellar gas column density distribution, CRs source distribution and energetics as well as the diffusion coefficient and Galactic magnetic fields. More details about those models are given in \cite{2015ApJ...799...86A}. Using the alternative Galactic diffuse emission models we repeated our $\sv$ upper limits calculation and found that those are affected at the few percent level for DM mass values smaller than 100 GeV, while no difference was apparent for larger DM masses. This is shown by the black dotted lines in Figure~\ref{fig:sigmav}.

\subsection{Forecast for future surveys}
\label{ssec:forcast}
Even though we obtain a relatively weak constraint on the DM cross section with our sample of 8 LSBGs, we expect fairly stringent constraints with a much larger sample.
As already mentioned in Sec.\ref{sec:data}, 781 LSBGs are found in the S16A HSC-wide footprint which roughly covers $200~{\rm deg^2}$. We estimate that 
${\cal O}(10^4)$ LSBGs will eventually be found within the $1400~{\rm deg^2}$ area of the HSC wide survey. 
In addition, future optical imaging surveys like LSST are expected to have better sensitivities which may make it possible to identify more nearby LSBGs. For example, the LSST will cover $\sim2\times 10^4~{\rm deg^2}$ of the sky with the limiting magnitude $\sim27$ mag in $g,r$ and $i$-band.
If the LSBGs are discovered by the LSST survey with the same rate of the HSC, we might be able to disentangle of order ${\cal O}(10^5)$ LSBGs in the near future.
\DvI{However, incorporating all those objects in an analysis of this kind is unrealistic because it is difficult to obtain accurate redshift measurements for all of them.} 
\OMvI{Conservatively, we assume that only a fraction of the soon-to-be-discovered LSBGs will have reliable distance measurements. In particular, we assume that future LSBG catalogs will have the same redshift measurement rate as that of the current HSC-LSBG catalog (8 objects within $\sim 200\; {\rm deg}^2$).}

\begin{figure}
 \begin{center}
   \includegraphics[width=12cm]{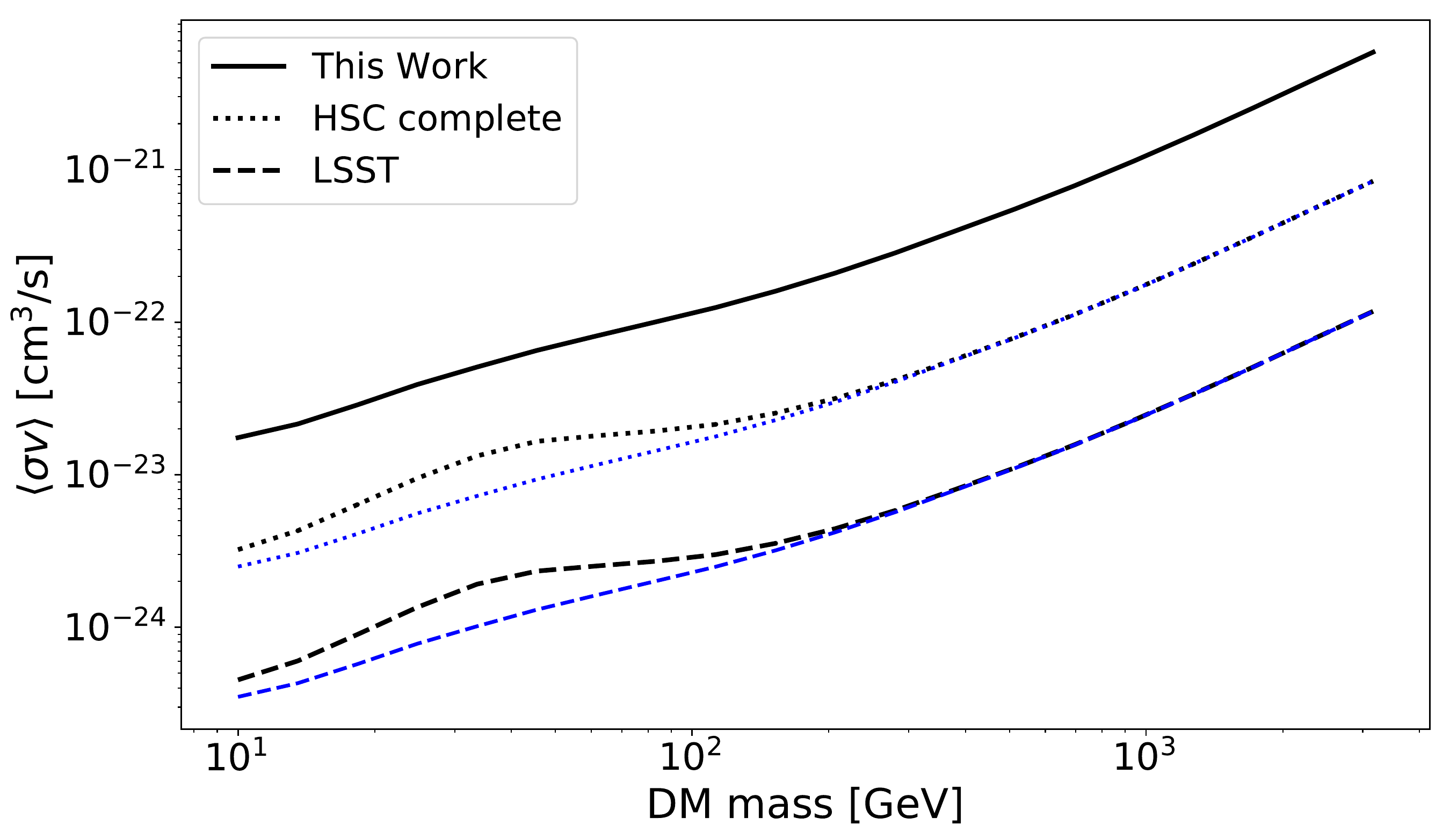}
 \end{center}
 \caption{
 \ajnvI{Expected} 95\% C.L. upper limits on the DM annihilation cross section in the $b\bar{b}$ channel. The solid black line shows the joint constraints obtained with the sample of 8 LSBGs in this work. \DvI{
 \ajnvI{Blue} dotted and dashed lines display 
 \ajnvI{the expected}
 constraints by assuming the HSC complete data and future LSST observations, respectively. 
 \ajnvI{Gray dotted and dashed }
 lines 
 \ajnvI{show}
 constraints 
 \ajnvI{obtained with a simple scaling with the number of expected LSBGs from our eight LSBGs.}
 }
 }
 \label{fig:sigmav_pros}
\end{figure}

\ajnvI{We further assume that future observations of  LSBGs will have a similar redshift and mass distribution to those in our current HSC catalog.}
\OMvI{Under this considerations, we estimate that the number of LSBGs to be detected is 60 and 400 for the ultimate HSC catalog and LSST, respectively.}
\DvI{In the case of the LSST forecast, we consider \ajnvI{only} half of the entire LSST footprint 
\ajnvI{to avoid the Galactic plane region, as this is a region that is highly contaminated by Galactic diffuse $\gamma$-ray emission.}
}
Figure \ref{fig:sigmav_pros} shows our predicted constraints on the DM cross section using the ultimate HSC catalog and forthcoming LSST objects. 

\DvI{In our forecast analysis we perform a joint analysis assuming that objects in each survey area have the same likelihood 
\ajnvI{as} 
our eight LSBGs.}
\DvI{In Figure~\ref{fig:sigmav_pros} we \ajnvI{show the expected constraints for the ultimate HSC and LSST surveys \OMvI{along with the} limits obtained with a simple scaling of the number of LSBGs. \OMvI{This scaling is done using the results of each survey area from the analysis of the current catalog of 8 LSBGs with measured redshifts.}}}

\OMvI{To understand the benefits of additional LSBG discoveries for DM searches it is useful to consider the \textit{signal fraction} $f$ defined~\cite{Charles:2016pgz} as
\begin{equation}\label{eq:signalfraction}
    f = \frac{n_{\rm sig}}{b_{\rm eff}},
\end{equation}
which approximately denotes the size of a DM signal ($n_{\rm sig}$ in units of photon counts) relative to the effective background photons ($b_{\rm eff}$). The quantity $f$ defined this way can be seen as a generalization of the signal-to-noise-ratio. In addition, the effective background can be estimated from the maximum likelihood fit covariance matrix (\textit{e.g.,} Eq.F.2 in Ref.~\cite{Charles:2016pgz}) and is given by 
\begin{equation}
b_{\rm eff} = \frac{N}{ \left( \sum_{k}\frac{P_{{\rm sig,}k}^{2}({\bf \mu})}{P_{{\rm bkg,}k}({\bf \theta})} \right) - 1},\label{eq:beff}
\end{equation}
where the index $k$ runs over all the energy bins and pixels in each ROI and  $N$ is the total number of observed photons in the corresponding ROI. The normalized signal and background model components can be straightforwardly computed as $P_{{\rm sig,}i}({\bf \mu}) = \lambda_{{\rm sig,}i}({\bf \mu}) / \sum_{k} \lambda_{{\rm sig,}k}({\bf \mu})$ and $P_{{\rm bkg,}i}({\bf \theta}) = \lambda_{{\rm bkg,}i}({\bf \theta}) / \sum_{k} \lambda_{{\rm bkg,}k}({\bf \theta})$, \OM{where $\lambda_{\rm sig}$ and $\lambda_{\rm bkg}$ represent the expected photon flux from a DM signal and the background flux, respectively. The signal parameters $\mu$ are relevant DM parameters like $\sv$ or $J$-factor. Equivalently, the nuisance parameter $\theta$ can be either the spectral normalization or  index of all the $\gamma$-ray background/foreground sources in the ROIs \D{(see more details in the \hyperref[sec:app]{Appendix})}.}
One noteworthy property of $b_{\rm eff}$ is that if the expected signal model and background model differ significantly such that $\sum_{k}P_{{\rm sig,}k}^{2}({\bf \mu})/P_{{\rm bkg,}k}({\bf \theta})\gg 1$ then $b_{\rm eff}$ will be proportionally less than $N$. In our case, this condition can be satisfied away from the Galactic plane due to the rapidly falling energy spectra of the background and smaller overall intensity. We also note that in the high energy regime the number of observed photons $N$ is drastically reduced thus making the search for DM in LSBGs \textit{signal limited targets}.}

\OMvI{As explained in a previous section, for a certain LSBG, the number of signal events $n_{\rm sig}$ is proportional to the product of the J-factor and $\sv$. By making the simple assumption that future LSBGs will have the same J-factors as those currently known, is easy to get that the total $n_{\rm sig}$ considered in the joint likelihood analysis is proportional to the number of LSBGs $ N_{\rm LSBG}$ times $\sv$. Upper limits at the 95\% CL on the total number of signal events $n_{\rm sig}$ are then obtained by increasing $n_{\rm sig}$ (\D{that is, increasing $\sv$}), while \OM{varying} all other parameters, until $-\Delta \log {\cal L}_{\rm stack}=2.0$ from the best-fit point.}


\ajnvI{We emphasise that it is essential to increase the number of spectroscopically confirmed LSBGs with future observations in order to obtain better and more efficient constraints on the DM annihilation cross sections using our proposed method.}

%
%

\ajnvI{We also note that the} 
forecast shown in Figure~\ref{fig:sigmav_pros}  
does not consider the actual distribution of the $J$-factor, which may affect the power of the predicted constraints.
As can be seen in Eq. (\ref{eq:j_factor}), the $J$-factor depends on the inverse square of the distance to the object but is only proportional to the DM halo mass.
This means that in case that future surveys find LSBGs 
closer to us than our current catalog of 8 LSBGs, then we might be able to obtain DM constraints that are stronger than those shown in the conservative predictions of Figure~\ref{fig:sigmav_pros}.


\section{Summary and Discussion}
\label{sec:summary}

For the first time, we have searched for dark matter emission in LSBGs using 8-years of observations of $\gamma$-ray data collected by \textit{Fermi}-LAT. No statistically significant signal was detected in these objects and thus we imposed new constraints on the DM annihilation cross-section. For this we used 8 LSBGs recently discovered by the HSC survey with reliable distance measurements.
The advantages of using LSBGs for constraining DM properties over other possible targets are: First, they are expected to have a low level of unresolved point point source contamination within. Secondly, they are typically 10 times more massive than \textit{e.g.} Milky Way dSphs. Lastly, the expected number of LSBGs will be very large (in fact much larger than that of dSphs). The main disadvantage of this novel dark matter target, compared to dSphs, is that they are much farther away and so they typically have smaller J-factors. However, we note that LSBGs are still located in the local Universe and the sensitivities can be incremented by increasing the detection of nearby LSBGs. Since the angular extensions of our LSBGs is sufficiently smaller than the Fermi PSF, in practice we were able to treat them as $\gamma$-ray point sources. This made  our analysis less sensitive to the model systematics and potential biases. 

We derived 95 \% C.L. upper limit on $\sv$ for each LSBG and also with a joint likelihood analysis using the full set of 8 objects. We found that $\sim 10^{-23}~[\rm cm^3/s]$ for a DM mass of 10 GeV self-annihilating into the $b\bar{b}$ channel. This constraint is $\sim 10^3$ times weaker than that obtained from Milky Way dSphs at the same DM mass (which is currently the most stringent).
We showed that a limiting factor in our methods corresponds to the low number of LSBGs considered. In the footprint of our S16A HSC data set, almost 800 LSBGs are identified. However, their distances are well measured for only the 8 objects used in this paper. Given the broad distance distribution of the entire sample, we presented a detailed systematic uncertainties analysis where we studied the impact of uncertainties in the distance measurement in our dark matter limits.



In the high energy limit where the number of detected photon counts is very low, the constraints on the annihilation cross-section scales with the number of objects $N$, instead of $\sqrt{N}$.
Indeed, above $\sim20$ GeV, \emph{Fermi}-LAT has rarely observed $\gamma$-ray photons in our ROIs. It follows that the number of objects is essential in obtaining constraints on the DM annihilation cross section. Although the current LSBG catalog used in this work contains $\sim800$ objects in the HSC $200~{\rm deg^2}$ footprint, this is going to increase to $\sim 6000$ in the final data release of HSC when the HSC survey finish to cover $1400~{\rm deg^2}$.
Moreover, in the near future, LSST will cover over $20000~{\rm deg^2}$ of sky with the depth of down to $27$ mag, and more than $10^5$ LSBGs can be expected to be identified. 
\DvI{We found that by means of conservative forecast analysis the expected constraints on the DM cross section using our method will not be stronger than those using 
\ajnvI{dSphs.}
}
\A{However, if the number of spectroscopically followed up LSBGs significantly increases, the constraint using those may reach to ones using dSphs because they are scaled with the number of detected LSBGs in the high DM mass.}
This is particularly the case in the high energy regime where the effective background is small compared to the expected signal counts.



\acknowledgments
We are grateful to Jenny E. Greene, Johnny P. Greco, Michael A. Strauss and Yoshiyuki Inoue for their careful read of the manuscript and useful comments.
This work is in part supported by MEXT Grant-in-Aid for Scientific Research on Innovative Areas (No.~15H05887, 15H05890, 15H05892, 15H05893, JP18H04340, 18H04359, 18J00277). 
The Hyper Suprime-Cam (HSC)
collaboration includes the astronomical
communities of Japan and Taiwan, and Princeton University.
The HSC instrumentation and software were developed by the National
Astronomical Observatory of Japan (NAOJ), the Kavli Institute for the
Physics and Mathematics of the Universe (Kavli IPMU), the University
of Tokyo, the High Energy Accelerator Research Organization (KEK), the
Academia Sinica Institute for Astronomy and Astrophysics in Taiwan
(ASIAA), and Princeton University.  Funding was contributed by the FIRST 
program from Japanese Cabinet Office, the Ministry of Education, Culture, 
Sports, Science and Technology (MEXT), the Japan Society for the 
Promotion of Science (JSPS),  Japan Science and Technology Agency 
(JST),  the Toray Science  Foundation, NAOJ, Kavli IPMU, KEK, ASIAA,  
and Princeton University. The Pan-STARRS1 Surveys (PS1) have been made possible through
contributions of the Institute for Astronomy, the University of
Hawaii, the Pan-STARRS Project Office, the Max-Planck Society and its
participating institutes, the Max Planck Institute for Astronomy,
Heidelberg and the Max Planck Institute for Extraterrestrial Physics,
Garching, The Johns Hopkins University, Durham University, the
University of Edinburgh, Queen's University Belfast, the
Harvard-Smithsonian Center for Astrophysics, the Las Cumbres
Observatory Global Telescope Network Incorporated, the National
Central University of Taiwan, the Space Telescope Science Institute,
the National Aeronautics and Space Administration under Grant
No. NNX08AR22G issued through the Planetary Science Division of the
NASA Science Mission Directorate, the National Science Foundation
under Grant No. AST-1238877, the University of Maryland, and Eotvos
Lorand University (ELTE). This paper makes use of software developed for the Large Synoptic
Survey Telescope. We thank the LSST Project for making their code
available as free software at http://dm.lsst.org.

\section*{Appendix}
\label{sec:app}
\begin{figure}
 \begin{center}
   \includegraphics[width=10cm]{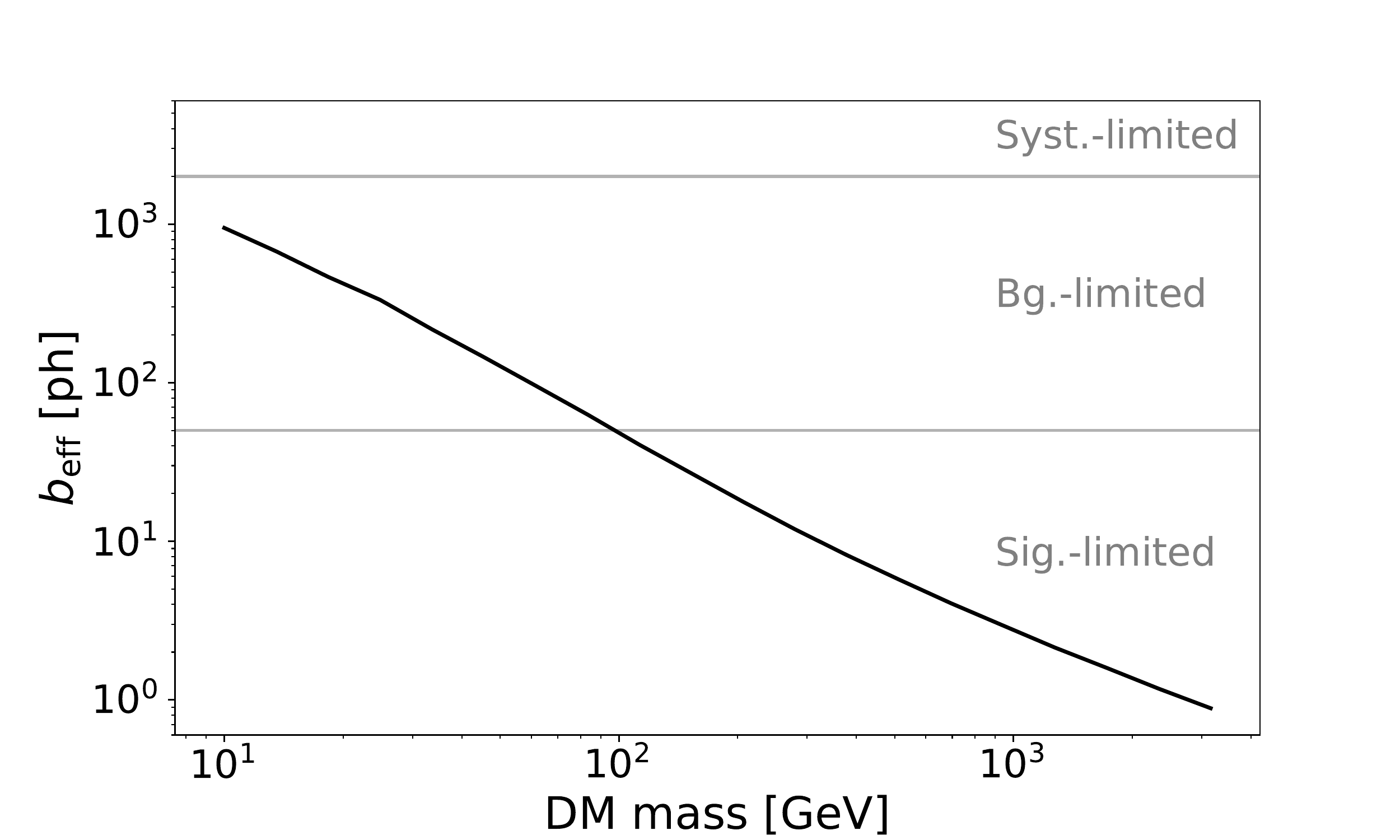}
 \end{center}
 \caption{
 The effective background for LSBG-285 (in units of photon counts) as a function of DM mass (GeV) for the $b\bar{b}$ channel. 
 Horizontal lines with $b_{\rm eff} = 2000$ and $b_{\rm eff} = 50$ represent cross-over levels between the systematics-limited and the background-limited, and the background-limited and signal-limited, respectively. 
 For more details see the text in \hyperref[sec:app]{Appendix}.}
 \label{fig:b_eff}
\end{figure}

\OM{In many cases, DM annihilation signals from objects of interest are expected to be small compared to the astrophysical background/foreground. Therefore, to discuss how the constraints on the annihilation cross-section are expected to increase with the number of objects included in the joint likelihood, we consider the useful definition of $b_{\rm eff}$ introduced in Eq.~\ref{eq:beff}. In practice, the quantities $\lambda_{\rm sig}$ and $\lambda_{\rm bkg}$ (necessary to compute $b_{\rm eff}$) can be computed with the use of the \texttt{gtmodel} tool within \texttt{FermiTools}. For $\lambda_{\rm sig}$, we computed infinite-statistics model maps, where the model only contained the DM target in question. As for $\lambda_{\rm bkg}$, we ran \texttt{gtmodel} assuming our best-fit background/foreground model obtained with our maximum-likelihood fitting procedure. Simple arithmetic operations follow to obtain $b_{\rm eff}$.}

Reference~\cite{2016PhR...636....1C} defines three different regimes in which the behaviour of the DM limits improve differently with additional data taking. First, in the systematics-limited regime ($b_{\rm eff}\gtrsim 2000$), the DM limits are expected to augment incrementally as our knowledge of the astrophysical background improves. DM limits obtained from targets in this regime are not expected to improve dramatically with additional data taking. Second, for objects in the background limited regime ($50\lesssim b_{\rm eff}\lesssim 2000$), the constraints can improve as $\propto \sqrt{N_{\rm obj}}$ (where $N_{\rm obj}$ is the number of objects). Third, for objects in the signal-limited regime ($b_{\rm eff}\lesssim 50$) the DM limits improve with $\propto N_{\rm obj}$. In Figure~\ref{fig:b_eff}, we show the effective background as a function of DM mass for one of the LSBGs in our study (LSBG-285). As can be seen, $b_{\rm eff}$ is in the signal-limited regime for $M_{\rm DM}\gtrsim 100$ GeV and is out of the systematic-limited regime for all DM masses.
Note that for lower masses ($M_{\rm DM}\lesssim 20$ GeV), the DM constraint is expected to scale only with $\sqrt{N_{\rm obj}}$. However, as seen in Figure~\ref{fig:sigmav_pros}, the constraint becomes more stringent. This seems to oppose the statement above but this is simply due to the lack of the statistics and the level of background is highly dependent on the sky position. The constraint will scale with $\sqrt{N_{\rm obj}}$ once we obtain a good amount of samples. On the contrary, in the signal-limited regime, as the background is subdominant, the fluctuation of the background on the sky position does not affect the constraints significantly. This is why we obtain a clean scaling relation of the DM constraint at high mass scale where the thermal relic scenario is still not excluded. 

\bibliography{bibdata}

\end{document}